\PassOptionsToPackage{table,dvipsnames}{xcolor}
\documentclass[sigconf,natbib=true,anonymous=false,screen=true]{acmart}

\usepackage{acronym}
\usepackage[noend]{algorithmic}
\usepackage{algorithm,balance}
\usepackage{bbm,acmart-taps}
\usepackage{beramono}
\usepackage{bm}
\usepackage{booktabs}
\usepackage[skip=0pt]{caption}
\usepackage[T1]{fontenc}
\usepackage{graphicx}
\usepackage{hyperref}
\usepackage{listings}
\usepackage{multirow}
\usepackage{natbib}
\usepackage{subcaption}
\usepackage{tabularx}
\usepackage{xcolor}
\usepackage{enumerate}
\usepackage[inline]{enumitem}
\usepackage{mleftright}
\usepackage{lipsum}
\usepackage[toc,page]{appendix}
\usepackage{centernot}
\usepackage{amsthm}

\copyrightyear{2023} 
\acmYear{2023} 
\setcopyright{cc} 
\acmConference[RecSys '23]{Seventeenth ACM Conference on Recommender Systems}{September 18--22, 2023}{Singapore, Singapore}
\acmBooktitle{Seventeenth ACM Conference on Recommender Systems (RecSys '23), September 18--22, 2023, Singapore, Singapore}\acmDOI{10.1145/3604915.3608788}
\acmISBN{979-8-4007-0241-9/23/09}

\usepackage{etoolbox}

\makeatletter

\definecolor{todo}{RGB}{200, 0, 0}
\definecolor{nk}{RGB}{0, 100, 0}
\definecolor{ho}{RGB}{0, 50, 150}
\definecolor{black}{gray}{0}
\definecolor{white}{gray}{1}

\acrodef{IR}{information retrieval}
\acrodef{LTR}{learning to rank}
\acrodef{RecSys}{recommender systems}
\acrodef{CTR}{click-through rate}
\acrodef{MF}{matrix factorization}
\acrodef{LBD}{learned beta distributions}
\acrodef{CMF}{Confidence-Aware Matrix Factorization}
\acrodef{MF}{Matrix Factorization}
\acrodef{PDF}{probability density function}
\acrodef{PMF}{probability mass function}
\acrodef{CDF}{cumulative distribution function}
\acrodef{RMSE}{root mean squared error}
\acrodef{MAE}{mean absolute error}
\acrodef{NDCG}{non discounted cumulative gain}
\acrodef{BPMF}{Bayesian Probabilistic Matrix Factorization}
\acrodef{LBD-S}{LBD Static}
\acrodef{LBD-A}{LBD Adaptive}

\theoremstyle{definition}

\author{Norman Knyazev}
\affiliation{%
	\institution{Radboud University}
	\city{Nijmegen}
	\country{The Netherlands}
}
\email{norman.knyazev@ru.nl}
\author{Harrie Oosterhuis}
\affiliation{%
	\institution{Radboud University}
	\city{Nijmegen}
	\country{The Netherlands}
}
\email{harrie.oosterhuis@ru.nl}

\acmSubmissionID{41}

\title{A Lightweight Method for Modeling Confidence in Recommendations with Learned Beta Distributions}
\begin{document}


\begin{abstract}
Most \ac{RecSys} do not provide an indication of confidence in their decisions.
Therefore, they do not distinguish between recommendations of which they are certain, and those where they are not.
Existing confidence methods for \ac{RecSys} are either inaccurate heuristics, conceptually complex or computationally very expensive.
Consequently, real-world \ac{RecSys} applications rarely adopt these methods, and thus, provide no confidence insights in their behavior.

In this work, we propose learned beta distributions (LBD) as a simple and practical recommendation method with an explicit measure of confidence.
Our main insight is that beta distributions predict user preferences as probability distributions that naturally model confidence on a closed interval, yet can be implemented with the minimal model-complexity.
Our results show that LBD maintains competitive accuracy to existing methods while also having a significantly stronger correlation between its accuracy and confidence.
Furthermore, LBD has higher performance when applied to a high-precision targeted recommendation task.

Our work thus shows that confidence in \ac{RecSys} is possible without sacrificing simplicity or accuracy, and without introducing heavy computational complexity.
Thereby, we hope it enables better insight into real-world \ac{RecSys} and opens the door for novel future applications.
\end{abstract}

\maketitle
\acresetall

\section{Introduction}

In order to best serve their users, \ac{RecSys} generally model user preferences to predict which item recommendations they would like~\citep{ricci2015recommender, steck2013evaluation}.
Generally, these models are constructed with collaborative filtering, where to predict the preference of one user-item combination, one considers what similar users thought of that item~\citep{marlin2004collaborative}.
The most fundamental of these methods is \ac{MF}~\citep{koren2009matrix} which represents users and items with embeddings, i.e., numerical vectors, where a high similarity between embeddings leads to a positive preference prediction, and vice versa.
Due to its effectiveness and simplicity, \ac{MF} is still one of the most competitive and widely-used \ac{RecSys} methods~\citep{rendle2019difficulty}.

In spite of being a heavily studied core \ac{RecSys} problem, the modeling of user preferences remains a very difficult task~\citep{ricci2015recommender}.
Part of this is due to the inherent uncertainty in user preferences: human behavior is complex and stochastic, and therefore, impossible to predict with perfect accuracy~\citep{anelli2020recsys, knijnenburg2012explaining}.
As a result, it is thus inevitable that even the best \ac{RecSys} models will make errors in their predictions.
However, despite these expected errors, the most commonly used \ac{RecSys} models do not give any sense of uncertainty in their predictions~\citep{wang2018ConfidenceAware, koren2011OrdRec}.
In other words, these models do not distinguish between predictions with high confidence, of which they are certain, or those with low confidence, of which they are unsure.
Consequently, current \ac{RecSys} give no insight into when recommendations are made with reliable confidence or doubtful guesswork~\citep{hullermeier2021aleatoric, klas2018uncertainty}.

Naturally, these shortcomings make the idea of confidence modeling for \ac{RecSys} very appealing.
A good measure of confidence could give insight into the behavior of a system, and give an indication of the extent of the error one can expect for a particular prediction.~\citep{rechkemmer2022confidence,mesas2020Exploiting}.
Furthermore, confidence enables functionalities such as exploration: recognizing which unexplored items have the potential of a high preference \citep{chu2011Contextual};
or conservative safety: guaranteeing a certain level of user experience by only following high-confidence preferences~\citep{moon2020confidence}.
In addition, uncertainty modeling could make \ac{RecSys} more explainable~\citep{zhang2020explainable}, and enable more insightful studies of user preferences.

In response, previous work proposed \ac{RecSys} with confidence by applying generative probabilistic models~\citep{salakhutdinov2008Bayesian, wang2018ConfidenceAware}.
To the best of our knowledge, all existing work assumes user ratings arise as samples from Gaussian probability distributions, where the ratings predicted by a \ac{MF} model provide the means of each distribution~\citep{mnih2007Probabilistic}.
\citet{salakhutdinov2008Bayesian} then apply a Bayesian approach and infer a probability distribution over embeddings that captures the uncertainty in the data.
But this inference has an impractically heavy computational cost, even when approximated, which seems to have prevented widespread real-world adoption.
As an alternative, \citet{wang2018ConfidenceAware} use static embeddings but vary the standard deviations of the Gaussians to model different levels of uncertainty.
However, this approach is very limited in the confidence patterns it can express, i.e. it only lets uncertainty vary on user and item levels separately and not over specific user-item combinations.
The lack of user-item interactions for confidence modeling is similarly a problem for \citet{koren2011OrdRec}, who instead capture the uncertainty by learning user-specific thresholds to assign each region of a Gaussian-like density to a single rating value.
The total density of each region then represents the probability of observing that rating.
While this allows confidence to vary, this solution is limited in the patterns it can express and does not appear easily interpretable.
As such, it appears the current \ac{RecSys} field has no method that effectively models uncertainty in an elegant and interpretable fashion, whilst also having practical computational requirements.

In this work, we introduce the \ac{LBD} as a simple and lightweight \ac{RecSys} method for a natural and explicit model measure of confidence.
Our method transforms a standard beta distribution into a discrete rating distribution, based on only two parameters that can be modeled with the same model-complexity as \ac{MF}.
By utilizing close approximations of their gradients~\citep{thompson1986Coulomb, robinson-cox1998Derivatives}, \acp{LBD} are optimized without high computational costs, thus making them very practical to work with.
Furthermore, \acp{LBD} can vary their confidence over different user-item combinations, with enough model expressiveness to capture complex patterns in uncertainty.
Our experimental results show \ac{LBD} has performance competitive with \ac{MF} and the confidence models of \citet{wang2018ConfidenceAware} and \citet{koren2011OrdRec}, while having a considerably stronger correlation between its confidence and its accuracy across ratings.

With the introduction of \ac{LBD} for \ac{RecSys} with confidence, we thus show that confidence modeling in \ac{RecSys} is possible without sacrificing accuracy or adding heavy computational complexity.
While our study is limited to \ac{MF},  the most foundational \ac{RecSys} model \citep{rendle2019difficulty}, the approach underlying \ac{LBD} is easily applicable or extendible to more complex \ac{RecSys} models.
For the \ac{RecSys} field, we hope our practical approach enables better insight into real-world recommendations, and stimulates future work into novel applications of uncertainty models for \ac{RecSys}.
\section{Related Work}
\label{sec:relatedwork}

Rating prediction is one of the core recommendation tasks in environments with significant amounts of explicit feedback \citep{adomavicius2007More,koren2009matrix,koren2011Advances,khan2021Deep}.
However, most widely-adopted models, such as \ac{MF} only focus on point predictions~\citep{shardanand1995Social,sedhain2015AutoRec,koren2009matrix,mnih2007Probabilistic}. As a consequence, they are also unable to differentiate between cases where the prediction task is easy and the prediction is likely correct, and those where the error is likely to be higher.

Being able to quantify confidence in individual predictions has many potential uses.
Showing confidence scores alongside recommendations is known to increase user satisfaction and platform trust~\citep{herlocker2000Explaining,mcnee2003Confidence,reilly2005Critiquing}.
\citet{mesas2020Exploiting} find that it may be possible to significantly increase precision by focusing only on the highest confidence predictions.
\citet{jeunen2021Pessimistic} and~\citet{adomavicius2007More} use lower confidence bounds instead of estimated relevance values to improve policy evaluation and recommender performance.
Conversely, upper confidence bounds can be used to recommend a more diverse range of items or to explore user preferences for items where they are poorly understood~\citep{bouneffouf2013RiskAwarea, chu2011Contextual}.

Existing confidence estimation methods can be broadly divided into three categories: heuristic, bootstrapping and generative approaches.
The algorithms in the first group extend point predictions with some predefined measures, for instance, the number of interactions for the user or the item \citep{mazurowski2013Estimating,adomavicius2007More,bernardis2019Estimating}.
However, whilst computationally light, the use of heuristics may also lead to poor generalization \citep{mazurowski2013Estimating,zhang2016Prediction}.
\citet{mazurowski2013Estimating} therefore proposes a bootstrapping approach that trains a recommender multiple times using different slices of the training data.
Uncertainty in a given rating is then represented by the variability of the prediction across different models.
However, training multiple models also means that this technique has a significant computational cost.

Generative methods instead view individual ratings as samples from a distribution, whose parameters have to be learned.
In practice, this distribution is generally assumed to be Gaussian with the same variance parameter for all ratings \citep{mnih2007Probabilistic,salakhutdinov2008Bayesian,lim2007Variational,wang2018ConfidenceAware,zhao2015Bayesian}.
However, further steps need to be taken to capture rating variability, as simple generative methods such as Probabilistic Matrix Factorization (PMF) have the same confidence for all ratings ~\citep{mnih2007Probabilistic}.
\ac{BPMF}~\citep{salakhutdinov2008Bayesian} extends PMF by integrating over all possible user and item vectors via Gibbs sampling~\citep{casella1992Explaining}.
The empirical sampling variance of the embeddings then represents the model's confidence in the correctness of the learned parameters.
However, this method of sampling is also expensive and the model thus difficult to apply in practice.
\citet{lim2007Variational} instead use variational methods for quicker convergence, albeit at the cost of accuracy and a worse fit to the data \citep{salakhutdinov2008Bayesian}.

Our approach is conceptually most similar to the following two approaches: \ac{CMF} of \citet{wang2018ConfidenceAware} and OrdRec of \citet{koren2011OrdRec}.
The former extends PMF by relaxing its assumption of shared variance.
Instead, each rating is assumed to be distributed with its own variance that is a product of learned global, user and item terms.
The authors also propose an analogous extension to \ac{BPMF}; however, the computational complexity arising from sampling still remains.
On the other hand, OrdRec~\citep{koren2011OrdRec} assumes a fixed-variance logistic distribution to model preference, but also learns a set of varying thresholds for each user.
The probability density falling between a pair of neighboring thresholds is then mapped to a single rating value and represents the probability of observing that rating.
This allows variance to vary per user, but does so in an oblique and hard to interpret manner.
We differ from existing work by relying on beta distributions to capture uncertainty, as we argue that this is a more appropriate choice than Gaussian or logistic distributions to reflect the variation in user ratings.

\section{Background: The Beta Distribution}
\label{sec:background_beta}
The beta distribution is a continuous probability distribution over the closed interval $[0, 1]$~\citep{johnson1994beta}.
Because it can easily be extended to any finite interval and can represent a great variety of distributions, the beta distribution has been widely applied to many different settings~\citep{gupta2004handbook}.
Despite this wide applicability, the distribution only has two parameters $\alpha>0$ and $\beta>0$. 
Its \ac{PDF} is defined as:
\begin{equation}
  f_B(x; \alpha, \beta) = \frac{x^{\alpha-1}(1-x)^{\beta-1}}{B(\alpha, \beta)},
\end{equation}
\noindent where the normalization factor, $B(\alpha, \beta)  = \int_{0}^{1} f_B(t;\alpha,\beta )dt$, is the complete regularized beta function that ensures that the \ac{PDF} integrates to one.
The mean of the distribution ($\mu$) is determined by the ratio of $\alpha$ and the sum $\alpha+\beta$, where an $\alpha$ higher than the $\beta$ means that the mean is closer to 1 than to 0, and vice versa.
The variance of the distribution depends on both the sum of the parameters as well as their product:
\begin{equation}
  \mathbb{E}[x] = \mu = \frac{\alpha}{\alpha + \beta},
  \qquad
  \mathbb{V}[x] = \frac{\alpha \beta}{(\alpha + \beta)^2 (\alpha + \beta + 1)}.
\end{equation}
We note that generally the beta distribution is non-symmetric, except when $\alpha = \beta$.
For instance, when $\alpha = \beta = 1$ it becomes equal to the (symmetric) uniform distribution.

The beta distribution models the likelihood function of the parameter underlying a Bernoulli trial, where $\alpha-1$ is the number of successes and $\beta-1$ the number of failures.
For an intuitive understanding, imagine $\alpha + \beta - 2$ flips of a biased coin where $\alpha - 1$ heads and $\beta - 1$ tails are observed.
If the coin had a probability $x$ of heads, the probability of the above result would be proportional to $x^{\alpha-1}(1-x)^{\beta-1}$.
We see that the PDF of the beta distribution is simply this probability for each $x$ value, but normalized to produce a valid distribution.

Inspired by this intuition, one can alternatively parameterize the beta distribution by its mean $\mu \in (0,1)$ and the sample size $\nu \in (0,\infty)$.
This corresponds to $\alpha = \nu\mu$ and $\beta = \nu(1-\mu)$ in the original parameterization, since $\mu = \alpha / (\alpha + \beta)$ and $\nu = \alpha + \beta$.
Intuitively for our likelihood example, $\mu$ would correspond to the observed mean and $\nu$ reflects the number of observations of the Bernoulli trial: thereby, $\nu$ can also be seen as a measure of confidence.

The \ac{CDF} of the beta distribution is not as straightforward:
\begin{equation}
	F_B(x; \alpha, \beta) = I_x(\alpha, \beta) = \frac{\int_{0}^{x} f_B(t)dt}{B(\alpha, \beta)},
\end{equation}
 where $I_x(\alpha, \beta)$ is the incomplete regularized beta function.
 Whilst no closed form for the CDF exists, it can be represented in terms of infinite power series or continued fractions \citep[p. 292, 385-389]{cuyt2008Handbook}.
Using these representations, close approximations of the CDF and its derivative w.r.t.\ $\alpha$ and $\beta$ can be computed efficiently~\citep{thompson1986Coulomb, robinson-cox1998Derivatives} and implementations of these approximations are already available in popular software packages such as Tensorflow~\citep{tensorflow2015} and Jax \citep{jax2018}.
 
Finally, while the beta distribution only covers the closed interval $[0,1]$, it can easily be used to create a distribution over any closed interval.
Let $R^\text{min}$ and $R^\text{max}$ indicate the endpoints of a closed interval for the random variable $x' \in [R^\text{min}, R^\text{max}]$.
We can transform the beta distribution to cover this interval by taking from it a sample $x\sim \text{beta}(\alpha, \beta)$ and transforming it: $x' = x(R^\text{max}-R^\text{min}) + R^\text{min}$.
Thus for each point $x'$ in the interval $[R^\text{min}, R^\text{max}]$, there is a corresponding $x$ in $[0,1]$: $x = \frac{x' - R^\text{min}}{R^\text{max}-R^\text{min}}$.
As a result, the PDF of the new distribution is simply:
$f_B'(x'; \alpha, \beta) = f_B(x; \alpha, \beta)/(R^\text{max} - R^\text{min})$,
and its CDF is:
$F_B'(x'; \alpha, \beta) =  F_B(x; \alpha, \beta)$.
The mean value undergoes the same transformation:
$\mathbb{E}[x'] = \mathbb{E}[x](R^\text{max}-R^\text{min}) + R^\text{min}$,
and the variance is multiplied by the size of the interval squared:
$\mathbb{V}[x'] = \mathbb{V}[x](R^\text{max}-R^\text{min})^2$.
Thereby, the beta distribution is not only useful to model ratios or probabilities, but also for modeling other values that are bound to other closed intervals.

\begin{figure*}
\centering
\resizebox{\textwidth}{!}{
    \includegraphics[width=\textwidth]{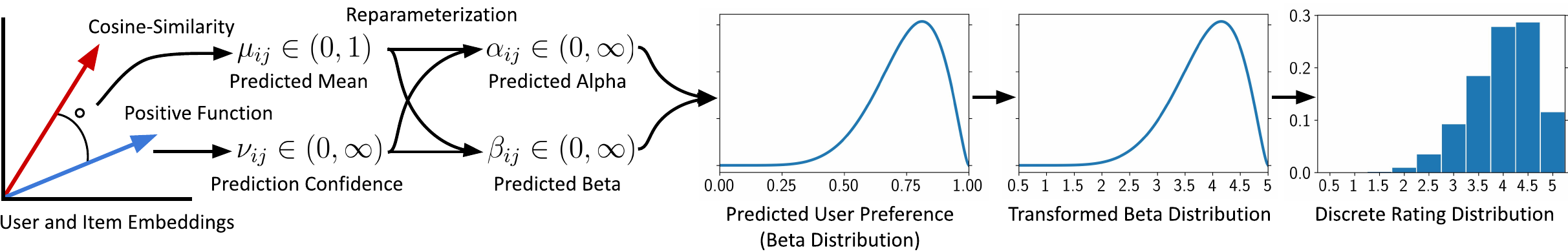}}
\caption{
Visualization of the \acf{LBD} method.
\ac{LBD} starts with user and item embeddings, their cosine similarity determines the predicted mean and a positive function its confidence.
These values are reparameterized to the input of a beta distribution, which models a prediction of user preference with explicit uncertainty.
This distribution is transformed and discretized to produce the predicted discrete rating distribution.
The entire \ac{LBD} method is differentiable and optimized with a cross-entropy loss.
}
\label{fig:overview}
\end{figure*}

\section{Method: Learned Beta Distributions for Rating Prediction}

Our goal is to introduce a rating prediction model that quantifies the confidence it has in its predictions without heavy computational costs.
In this section, we propose \acl{LBD} as an expressive and conceptually simple method for rating prediction with an elegant measure of confidence.
The remainder of this section is structured as follows:
First, we detail how a beta distribution can be transformed into an optimizable distribution over discrete ratings (Section~\ref{subsec:discrete}).
Second, several approaches are proposed to model the parameters of a unique rating distribution for each user-item combination (Section~\ref{subsec:modeling_mean_confidence}).
Third, two alternatives to modeling user and item biases into the parameters of the model are introduced (Section~\ref{subsec:biases}).
Fourth, static and adaptive binning strategies for discretization are proposed (Section~\ref{subsec:binning}).
Fifth, a summary of the complete approach is laid out, starting from user and item embeddings and ending in discrete rating distributions (Section~\ref{subsec:overview}).
Finally, we compare \ac{LBD} with the existing \ac{MF}, \ac{CMF} and OrdRec approaches in conceptual terms (Section~\ref{subsec:comparison}).

\subsection{From a beta distribution to an optimizable discrete rating distribution}
\label{subsec:discrete}

In this subsection, we describe how a beta distribution can be transformed to model a discrete rating distribution, and subsequently optimized with a cross-entropy loss.

We assume that users can provide ratings on a discrete and linear scale:
$R \in \{R_1, R_2, \ldots, R_n\}$, where $n$ is the number of possible rating values and the difference between $R_1$ and $R_2$ is equal to that between $R_2$ and $R_3$, etc.
To denote the minimum and maximum values, we use $R_1 = R^\text{min}$ and $R_n = R^\text{max}$, respectively.

Let $\alpha_{ij}$ and $\beta_{ij}$ be the parameters for the distribution of rating $R_{ij}$, corresponding with user $i$ and item $j$.
We begin by transforming the associated beta distribution to the interval spanning the range of possible rating values: $[R^\text{min}, R^\text{max}]$.
As discussed in Section~\ref{sec:background_beta}, a sample from a beta distribution:
$x \sim \text{beta}(\alpha_{ij}, \beta_{ij})$
can be transformed accordingly:
$
x' = x(R^\text{max}-R^\text{min}) + R^\text{min}.
$%
The resulting distribution over $x'$ has the same shape as the beta distribution, but is transposed and stretched over the rating interval.
The PDF is easily mapped back to the original distribution:
$f_B'(x'; \alpha, \beta) = f_B(x; \alpha, \beta) / (R^\text{max}-R^\text{min})$
and the same goes for the CDF:
$F_B'(x'; \alpha, \beta) =  F_B(x; \alpha, \beta)$.

While our new distribution is on the correct interval, it still has to be discretized.
This can be done straightforwardly by splitting the interval into bins, each representing the probability mass of one rating value.
Let the $W_{ij,r} \in [0,1]$ parameter represent the bin width for rating $r$ s.t.\ $\sum_{r=1}^n W_{ij,r} = 1$. The lower edge of rating $R_{r}$ can then be defined for $r > 1$ as
$E_{r} = E_{r-1} + (R^\text{max}-R^\text{min})W_{ij,r-1}$, and $E_{1}=R^\text{min}$ for the first edge.
The probability of a rating is then simply the area under the PDF and inside the corresponding bin.
We note that an equivalent discretization can be achieved by binning the original beta distribution, the bin edges here are simply transformed inversely: $e_{r} = (E_{r}- R^\text{min})/(R^\text{max}-R^\text{min})$.
Therefore, it is also straightforward to express the probabilities in terms of the CDF of the underlying beta distribution: 
\begin{align}
P\big(R_r ; \alpha_{ij}, \beta_{ij} \big)
&= F_B'\big(E_{r+1} ; \alpha_{ij}, \beta_{ij} \big) - F_B'\big(E_{r} ; \alpha_{ij}, \beta_{ij} \big)\nonumber\\
 &= F_B\big(e_{r+1} ; \alpha_{ij}, \beta_{ij} \big) - F_B\big(e_{r} ; \alpha_{ij}, \beta_{ij} \big).
\end{align}

Importantly, our transformation from the beta CDF ($F_B$) to the rating probability $P\big(R_r ; \alpha_{ij}, \beta_{ij}\big)$ is completely differentiable.
Furthermore, the gradients w.r.t.\ $\alpha_{ij}$ and $\beta_{ij}$ can be computed using the close approximations of previous work~\citep{thompson1986Coulomb, robinson-cox1998Derivatives}.
In other words, our discrete rating probability is fully differentiable to its input parameters.
As a result, we can easily use it to optimize a cross-entropy loss:
\begin{equation}
\mathcal{L}(\alpha, \beta) = \frac{1}{Z} \sum_{i} \sum_{j} \log P(R_{ij} ; \alpha_{ij}, \beta_{ij}).
\end{equation}

Thus, we have shown that beta distributions can be transformed into discrete rating distributions in a straightforward and elegant manner.
Since the result is fully differentiable, it is easily optimizable; in particular, we propose to maximize the log-likelihood of observed rating data by minimizing a cross-entropy loss.

\subsection{Modeling means and confidences}
\label{subsec:modeling_mean_confidence}

We have introduced a method for producing a discrete rating distribution for a rating $R_{ij}$ from parameters $\alpha_{ij}$, $\beta_{ij}$ and $W_{ij}$.
In this subsection, we propose a simple way of modeling $\alpha_{ij}$ and $\beta_{ij}$ based on the interactions between user and item embeddings.

To start, we make use of the alternative parameterization described in Section~\ref{sec:background_beta} and introduce the mean $\mu_{ij} \in (0,1)$ and confidence $\nu_{ij} \in(0,\infty)$.
These map back to the original parameterization through:
$\alpha_{ij} = \mu_{ij}\nu_{ij}$ and $\beta_{ij} = (1-\mu_{ij})\nu_{ij}$.
Furthermore, inspired by the simplicity of \ac{MF}, we introduce a non-zero learned embedding for each user $U_i \in \mathbb{R}^D$  and for each item $V_j \in \mathbb{R}^D$.
Our first insight is that because $\mu_{ij} \in (0,1)$, it can be modeled elegantly with the cosine similarity:
\begin{equation}
	\label{eq:core_model}
	\mu_{ij} = \dfrac{1}{2}+\dfrac{1}{2}\text{Cosine-similarity}(U_i, V_j) = \dfrac{1}{2}+\dfrac{1}{2}\dfrac{U_i^\intercal V_j}{||U_i||\,||V_j||}.
\end{equation}
Thereby, a higher similarity between the directions of the embeddings results in a higher mean of the predicted rating distribution.
Our second insight is that any differentiable transformation from the embeddings to a positive scalar can be used for the confidence parameter.
Whilst many transformations are possible, we propose the following three:
\begin{equation}
\nu_{ij}^{\text{norm}} = ||U_i||\,||V_j||,
\quad
\nu_{ij}^\text{sum} = ||U_i + V_j||,
\quad
\nu_{ij}^\text{dot} = |U_i^\intercal V_j|.
\end{equation}

With $\nu_{ij}^{\text{norm}}$, the confidence parameter is equal to the product of the $L_2$ norms of the embeddings.
Thus users or items with larger embeddings will result in more confident predictions.
Alternatively, one could interpret this as each user and item having a unique confidence term, expressed in the distance of their embedding vector from the origin.
The simplicity of $\nu_{ij}^{\text{norm}}$ makes it easy to understand, but it therefore also lacks expressiveness and cannot capture complex patterns that vary over specific user-item interactions.

The $L_2$ norm of the summation of the embeddings is used to model confidence in $\nu_{ij}^\text{sum}$.
Here both the direction and the distance of the embeddings determine the resulting confidence.
As a result, this model can capture more complex interactions between users and items, but is also intuitively more difficult to understand.

Lastly, $\nu_{ij}^\text{dot}$ uses the absolute value of the dot product of the embeddings.
Thus, confidence is also determined by both the direction and the distance of the embedding vectors.
However, because the absolute value is used, both high similarity and dissimilarity result in high confidence predictions.
The dot product highly correlates with the cosine similarity, which could also mean there is a high correlation between $\mu_{ij}$ and $\nu_{ij}^\text{dot}$.
It is unclear whether such a correlation is desirable in practice.

To summarize, we have introduced a novel approach to modeling a beta distribution by basing its input parameters on the interaction between user and item embeddings.
In particular, we propose to model mean ratings by cosine similarity and the confidence by one of several possible transformations.
Thereby, our approach has the same model-complexity as \ac{MF}, since both only use user and item embeddings.
But whereas \ac{MF} only uses them to model pointwise ratings, \ac{LBD} models predictions as probability distributions.

\subsection{Modeling user and item biases}
\label{subsec:biases}

An important feature of standard \ac{MF} models is that they can also include bias weights per user and item~\citep{koren2009matrix,koren2009Collaborative,johnson2014logistic}.
These bias weights can increase all the ratings of a user or an item, for instance, to capture that certain users tend to give higher ratings than the average user.
Because adding these weights is very effective for \ac{MF}, we propose adding them to our \ac{LBD} model as well.

The addition of bias weights to the beta distribution is straightforward. We introduce the optimizable scalars $a \in \mathbb{R}$ and $b \in \mathbb{R}$, and denote $a_0$ for a global weight, $a_i$ for a user weight and $a_j$ for an item weight, and the analogous weights $b_0$, $b_i$ and $b_j$.
These bias weights are then simply added to the weights predicted with the embeddings:
\begin{equation}
\alpha_{ij}' =  \max(a_0 +  a_{i} + a_{j} + \alpha_{ij}, \epsilon),
\quad
\beta_{ij}' = \max(b_0 + b_{i} + b_{j} + \beta_{ij}, \epsilon),
\end{equation}
where $\epsilon$ is a very small positive quantity to ensure that $\alpha_{ij}' > 0$ and $\beta_{ij}' > 0$.

While the addition may appear simple, it also has a Bayesian interpretation, as the normalized product of beta distributions can also be produced by summing their parameters accordingly:
\begin{align}
 f_B(a_0+1, b_0+1) \,
  &f_B(a_i+1, b_i+1)\,
  f_B(a_j+1, b_j+1)\,
  f_B(\alpha_{ij}, \beta_{ij})\nonumber\\
  &\propto f_B(a_0 + a_i + a_j + \alpha_{ij}, b_0 + b_i + b_j + \beta_{ij}).
\end{align}
Therefore, we can interpret $f_B(a_0 + 1, b_0 + 1)$ as a global prior distribution, $f_B(a_i+1, b_i+1)$ as the likelihood function given that user $i$ is involved, the same for $f_B(a_j+1, b_j+1)$ and item $j$, and finally, $f_B(\alpha_{ij}, \beta_{ij})$ as the likelihood function for the interaction between $i$ and $j$.
While this Bayesian interpretation is valid, we note that our optimization only considers the final distribution: therefore, it does not directly optimize the accuracy of the prior and likelihood decomposition.

Additionally, we also propose bias weights in terms of the alternative parameterization.
We introduce the optimizable scalars $u \in (0,1)$ and $v \in (0,\infty)$, where $u_0$ and $v_0$ represent global weights and $u_i$, $u_j$, $v_i$ and $v_j$ user and item specific weights.
The weights $v$ are used as multipliers on the confidences $\nu$: thus higher $v$ result in more confident predictions, and vice versa.
For the $u$ weights, we create a piecewise linear mapping that places $\mu_{ij}' = 0.5$ when $\mu_{ij} = u_0 u_i u_j$: thus high $u$ values result in a lower predicted mean, and vice versa.
Formally, the bias weights are applied as follows:
\aptLtoX[graphic=no,type=html]{
\begin{gather}
\begin{aligned}
	\label{eq:mu_adjustment}
	\nu_{ij}' = v_0 \, v_i \, v_j \, \nu_{ij}, \quad
	\mu_{ij}' = 
		\begin{cases}
			\dfrac{\mu_{ij}}{2\,u_0 \, u_i \, u_j} &\text{ if } \mu_{ij} \geq u_0 \, u_i \, u_j,\\
			\dfrac{1}{2} + \dfrac{\mu_{ij}-u_0 \, u_i \, u_j}{2(1- u_0 \, u_i \, u_j)} &\text{ otherwise.}
		\end{cases} 
	\end{aligned}
\end{gather}
}{ \begin{gather}
\begin{aligned}
	\label{eq:mu_adjustment}
	\nu_{ij}' = v_0 \, v_i \, v_j \, \nu_{ij}, \quad
	\mu_{ij}' = 
		\begin{cases}
			\dfrac{\mu_{ij}}{2\,u_0 \, u_i \, u_j} &\text{ if } \mu_{ij} \geq u_0 \, u_i \, u_j,\\
			\dfrac{1}{2} + \dfrac{\mu_{ij}-u_0 \, u_i \, u_j}{2(1- u_0 \, u_i \, u_j)} &\text{ otherwise.}
		\end{cases} 
		\raisetag{16.5pt}
	\end{aligned}
\end{gather} }
Since our proposed applications of bias weights are fully differentiable, they effortlessly integrate into our \ac{LBD} model.

\subsection{Modeling static and adaptive discretization strategies}
\label{subsec:binning}

We have described how user and item embeddings and biases can model the parameters of a beta distribution.
As discussed in Section~\ref{subsec:discrete}, \ac{LBD} creates a discrete rating distribution through binning according to the bin width  parameters $W$.
We propose a static binning strategy for modeling $W$: LBD-S; and an adaptive binning strategy: LBD-A.
\acused{LBD-S}
\acused{LBD-A}

The static strategy of LBD-S straightforwardly makes each bin equisized: $\forall i,j,r: W_{ij,r} = \frac{1}{n}$.
In terms of model-complexity, this is the most efficient choice as none of the $W$ parameters have to be learned.
Implicitly this strategy assumes a linear relation between preference and ratings, i.e., the uniform distribution with $\alpha = 1$ and $\beta = 1$ will translate to a uniform discrete distribution.
However, this assumption may not be correct, e.g., in the Movielens dataset~\citep{harper2015movielens}, the whole ratings ($\{1,2,3,4,5\}$) are much more frequent than the half ratings ($\{0.5,1.5,2.5,3.5,4.5\}$).
This difference in frequency seems to be an artifact of the rating interface, which leads users to choose whole-number ratings over the others, and is unlikely to reflect an actual clustering in the underlying preferences~\citep{peska2022Effect, jin2003Collaborative}.
Importantly, such patterns are difficult to capture for LBD-S, since it assumes that probability mass is smoothly spread over all rating values.

In contrast, the adaptive binning strategy varies the bin widths per rating, user and item, and thereby, it can better capture frequency differences between certain rating values.
Inspired by the discretization of \citet{koren2011OrdRec}, we introduce the unnormalized bin widths $w_{ij,r} = \exp(\theta_i^r + \theta_j^r)$ with the optimizable parameters $\theta_i^r$ and $\theta_j^r$ for each user $i$, item $j$ and rating value $r$.
The bin widths are then simply obtained by normalizing:  $W_{ij,r} = w_{ij,r} / \sum_{r'=1}^n w_{ij,r'}$.
By increasing the bin width of certain rating values relative to others, these rating values can become more probable without changing the underlying beta distribution.
Thereby, LBD-A can capture non-smooth patterns that LBD-S cannot, e.g., that half ratings are less probable than whole ratings, or whether a certain user is more likely to translate a specific level of preference to a higher rating than most other users~\citep{jin2003Collaborative}.
Importantly, the increased expressiveness of LBD-A over LBD-S comes at the cost of only a few additional parameters for each user and item.

\subsection{Overview of the \ac{LBD} method}
\label{subsec:overview}

Now that all the components of the \acf{LBD} method have been introduced, we summarize how these components come together to form a single approach.
The visualization in Figure~\ref{fig:overview} accompanies this subsection.

\begin{figure*}

\setlength{\tabcolsep}{0.015cm}
\centering
\begin{tabular}{cccc}
    \hspace{0.27cm}Embeddings
    & Item 1
    & Item 2
    & Item 3
    \\
    \includegraphics[scale=0.24, trim=0 -0.55cm 0 0, clip]{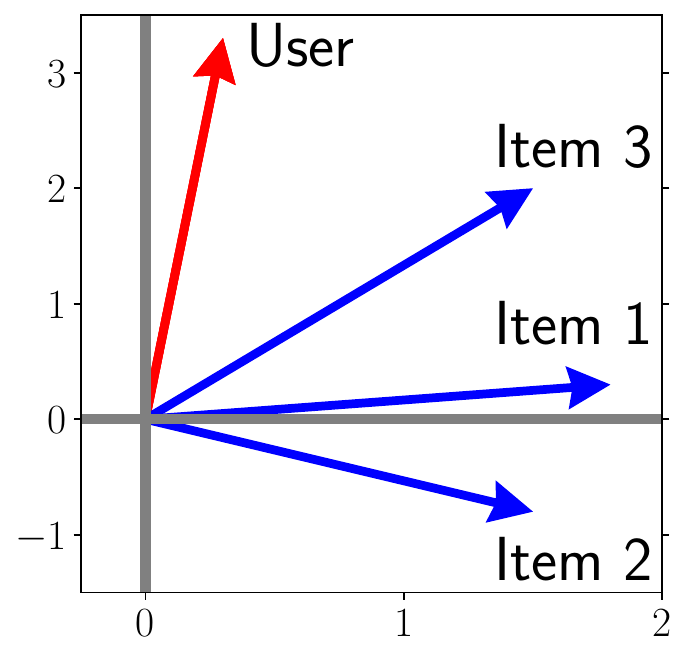}
    & \includegraphics[scale=0.43]{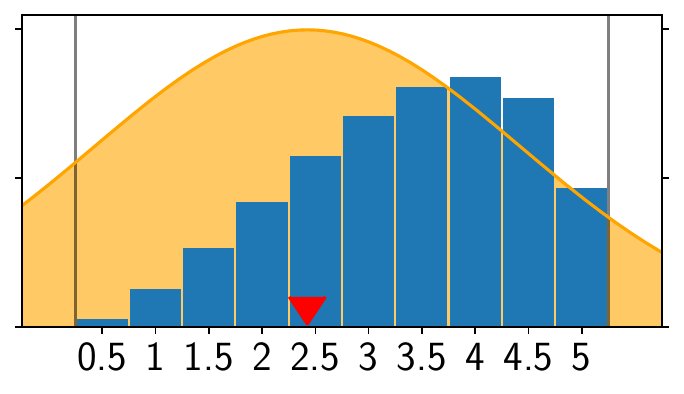}
    & \includegraphics[scale=0.43]{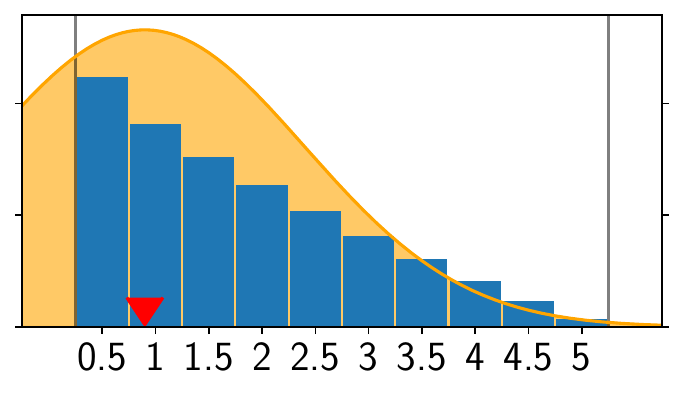}
    & \includegraphics[scale=0.43]{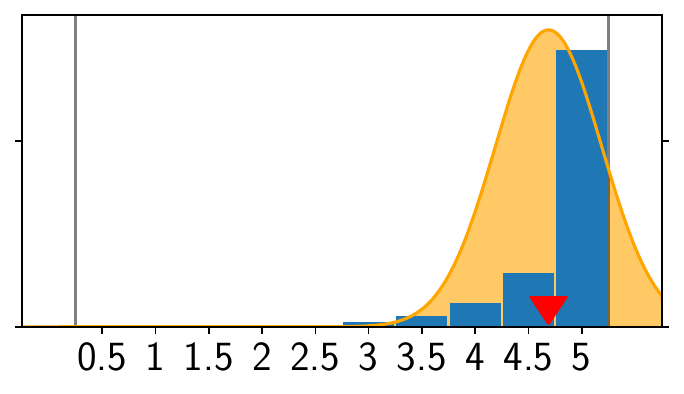}
\end{tabular}
\caption{
Example comparison of how \ac{MF}, \ac{CMF} and \ac{LBD} make predictions for a user and three different items.
The prediction of \ac{MF} is pointwise (red triangle);
that of \ac{CMF} is a Gaussian distribution (orange continuous probability density function);
and that of \ac{LBD} is a discrete rating prediction based on an underlying beta distribution (blue discrete probability mass function).
}
\label{fig:comparison}
\end{figure*}

Similar to \ac{MF}, \ac{LBD} has an optimizable embedding vector for each user and item, the cosine similarity between these vectors determines the mean prediction ($\mu$), and a positive function is used to get a positive scalar indicating the confidence in the prediction ($\nu$).
The mean and confidence are then mapped to $\alpha$ and $\beta$ values and bias terms are also added. The resulting values serve as input parameters to a beta distribution.
This beta distribution is the prediction of user preferences with explicit uncertainty; preferences are modeled as ratios in the interval $[0,1]$ and the beta \ac{PDF} ($f_B$) indicates how probable each possible value is predicted to be.
This distribution is then transformed to match the range of possible rating values and discretized to give exact probabilities per rating. 
Discretization can be done either using equally sized bins or by adjusting each region's width $w_{ij,r}$ based on the user and item terms ($\theta$).
Because each step between the embeddings and the final discrete rating distribution is differentiable, one can compute the gradient of a rating probability w.r.t.\ the underlying embeddings.
We utilize this differentiability to optimize the model with a cross-entropy loss, so that the predicted discrete rating probabilities maximize the likelihood of observed rating data.

Thereby, our \ac{LBD} method predicts user ratings with explicit uncertainty in its predictions.
Importantly, it does this with the same model-complexity as \ac{MF}, and unlike previous methods that utilize Gaussian distributions~\citep{wang2018ConfidenceAware, salakhutdinov2008Bayesian}, \ac{LBD} can have non-symmetric variance that can capture user-item interactions, never predicts ratings outside of the valid rating range, can be described using just two parameters and does not involve heavy computational costs in its optimization and prediction.

\subsection{Comparison with existing methods}
\label{subsec:comparison}

Before \ac{LBD} is compared with the existing \ac{MF}~\citep{koren2009matrix}, \ac{CMF}~\citep{wang2018ConfidenceAware} and OrdRec~\citep{koren2011OrdRec} methods in experimental terms in Section~\ref{sec:results}, we first consider key conceptual differences between the methods.
To accompany the discussion, Figure~\ref{fig:comparison} visualizes the predictions of \ac{MF}, \ac{CMF} and \ac{LBD} for an example scenario with one user and three items.

To start, \ac{MF} differs from \ac{CMF}, OrdRec and \ac{LBD} in that its pointwise predictions are without any sense of confidence, and accordingly, they are presented as points on the x-axis.
As a result, while \ac{CMF} and \ac{LBD} show high uncertainty about their predictions for item 1 and 2, \ac{MF} gives no indication that it may be less reliable there.

Next, we compare \ac{CMF} and \ac{LBD}, as both give predictions in the form of a distribution: \ac{CMF} as a Gaussian distribution and \ac{LBD} as a discrete probability distribution based on an underlying beta distribution.
We argue that the usage of a beta distribution by \ac{LBD} provides three key advantages over Gaussians for rating prediction:
\aptLtoX[graphic=no,type=html]{
(i) Beta distributions can be bounded to only give probability inside the possible range of ratings, whereas Gaussians are unbounded.
This problem is illustrated across all items in Figure~\ref{fig:comparison} as the \ac{PDF} of \ac{CMF} gives much probability outside the valid range of ratings.
(ii) Beta distributions can have non-symmetric variance, while Gaussians cannot.
For instance, \ac{LBD} thinks the lowest rating of item 2 is most probable, and puts the remainder of its probability on the higher ratings.
In contrast, \ac{CMF} also thinks a low rating is most likely, but has to give equal probability to ratings below and above it, including ratings out of the valid range.
(iii)
Lastly, \ac{LBD} can be fully parameterized with minimal model-complexity, since it only requires an interaction between two embeddings to compute all its parameters.
Conversely, this is not possible for a Gaussian because its mean is unbounded, and as a result, \ac{CMF} needs additional parameters for its confidence model (not visualized in Figure~\ref{fig:comparison}).
 }{  \begin{enumerate*}[label=(\roman*)]
\item Beta distributions can be bounded to only give probability inside the possible range of ratings, whereas Gaussians are unbounded.
This problem is illustrated across all items in Figure~\ref{fig:comparison} as the \ac{PDF} of \ac{CMF} gives much probability outside the valid range of ratings.
\item Beta distributions can have non-symmetric variance, while Gaussians cannot.
For instance, \ac{LBD} thinks the lowest rating of item 2 is most probable, and puts the remainder of its probability on the higher ratings.
In contrast, \ac{CMF} also thinks a low rating is most likely, but has to give equal probability to ratings below and above it, including ratings out of the valid range.
\item
Lastly, \ac{LBD} can be fully parameterized with minimal model-complexity, since it only requires an interaction between two embeddings to compute all its parameters.
Conversely, this is not possible for a Gaussian because its mean is unbounded, and as a result, \ac{CMF} needs additional parameters for its confidence model (not visualized in Figure~\ref{fig:comparison}).
\end{enumerate*} }

Finally, our visualization in Figure~\ref{fig:comparison} does not include OrdRec because its differentiating characteristics are difficult to display.
OrdRec uses the logistic distribution, which is very similar to the Gaussian distribution but with heavier tails.
However, unlike \ac{CMF} and similar to \ac{LBD}, it also applies discretization through the same adaptive binning strategy as described in Section~\ref{subsec:binning}.
Thus unlike \ac{CMF}, OrdRec can transform the shape of its rating distribution through adaptive binning, enabling it to have asymmetric variance.
Nevertheless, the adaptive binning strategy is relatively limited in its expressiveness (cf. $w_{ij,r}$ and $\nu_{ij}$).
Furthermore, it is also unclear whether the symmetric logistic function, with its infinite coverage and heavy tails, is appropriate to model user preferences.
In contrast, \ac{LBD} captures both a measure of preference and confidence within its embeddings, and can already express asymmetric preference distributions before discretization is applied.
The remainder of this paper will investigate whether the theoretical advantages of \ac{LBD} can be verified experimentally and lead to more effective modeling of uncertainty.

\section{Experimental Setup}

The experiments carried out for this paper are aimed at answering the following four research questions:
\aptLtoX[graphic=no,type=html]{
\begin{enumerate}
    \item[\textbf{RQ1}] What modeling of the parameters of \ac{LBD} results in the highest predictive performance for rating prediction?
    \label{rq:parameter}
    \item[\textbf{RQ2}] Does our \ac{LBD} approach provide rating prediction performance that is competitive with \ac{MF}, \ac{CMF} and OrdRec?
    \label{rq:performance}
    \item[\textbf{RQ3}] Is there a stronger correlation between the confidence and accuracy of predictions by \ac{LBD} than for \ac{CMF} and OrdRec?
    \label{rq:confidence}
    \item[\textbf{RQ4}] Does the confidence modeling of \ac{LBD} translate to higher performance compared to  \ac{MF}, \ac{CMF} and OrdRec in a high-precision targeted recommendation task?
    \label{rq:downstream}
\end{enumerate} }{ \begin{enumerate}[label=\textbf{RQ\arabic*}]
	\item What modeling of the parameters of \ac{LBD} results in the highest predictive performance for rating prediction?
	\label{rq:parameter}
	\item Does our \ac{LBD} approach provide rating prediction performance that is competitive with \ac{MF}, \ac{CMF} and OrdRec?
	\label{rq:performance}
	\item Is there a stronger correlation between the confidence and accuracy of predictions by \ac{LBD} than for \ac{CMF} and OrdRec?
	\label{rq:confidence}
	\item Does the confidence modeling of \ac{LBD} translate to higher performance compared to  \ac{MF}, \ac{CMF} and OrdRec in a high-precision targeted recommendation task?
	\label{rq:downstream}
\end{enumerate} }
Correspondingly, the goals behind these research question are:
\aptLtoX[graphic=no,type=html]{
(i) to find out what version of \ac{LBD} is the most accurate at rating prediction;
(ii) to evaluate whether the confidence modeling of \ac{LBD} sacrifices competitive accuracy;
(iii) to verify whether \ac{LBD} has a better indication of confidence than the existing confidence-aware methods; and
(iv) to investigate how the confidence modeling of \ac{LBD} could be beneficially leveraged in a practical downstream task.
 }{ \begin{enumerate*}[label=(\roman*)]
\item to find out what version of \ac{LBD} is the most accurate at rating prediction;
\item to evaluate whether the confidence modeling of \ac{LBD} sacrifices competitive accuracy;
\item to verify whether \ac{LBD} has a better indication of confidence than the existing confidence-aware methods; and
\item to investigate how the confidence modeling of \ac{LBD} could be beneficially leveraged in a practical downstream task.
\end{enumerate*} }

\subsection{Dataset}
We base our experiments on the MovieLens 10M dataset~\citep{harper2015movielens}, %
where ratings from $69878$ users are spread over $10677$ items on a $0.5$ point scale between $0.5$ and $5$.
In order to perform cross-validation, we randomly split the user ratings into $10$ folds.
Each experiment is repeated $10$ times, so that each fold is used for evaluation once, while the remaining $9$ folds are further split into $95\%$ train and $5\%$ validation data.
To avoid irrelevant cold-start problems in evaluation, we further ensure that the evaluation set never contains ratings involving users or items that are not in the training set.

\subsection{Baselines, metrics and parameter tuning}

We compare our \ac{LBD} method with three baselines of similar model-complexity: \ac{MF}~\citep{koren2009matrix,mnih2007Probabilistic}, \ac{CMF}~\citep{wang2018ConfidenceAware} and OrdRec~\citep{koren2011OrdRec}.
\ac{MF} is a foundational \ac{RecSys} method and remains a very strong rating prediction baseline when tuned well~\citep{rendle2019difficulty,rendle2020Neural}.
Furthermore, it does not have any confidence modeling, and thereby, allows us to evaluate whether the confidence modeling of \ac{LBD} prevents it from providing competitive rating prediction accuracy.
\ac{CMF} performs confidence modeling through Gaussian distributions with varying variance. %
Lastly, OrdRec models confidence by partitioning a logistic distribution into non-overlapping regions,  each representing the probability mass assigned to a possible rating value.

For the comparison with the baselines, we use the two best \ac{LBD} parameterizations identified for \ref{rq:parameter}.
Both use $\alpha$ and $\beta$ bias terms  with embeddings of size $512$ to jointly model preference and confidence, where the latter is represented via $\nu^\text{sum}$.
We consider both static binning (\acs{LBD-S}) and adaptive binning (\acs{LBD-A}).

To evaluate various aspects of recommendation performance, we use a variety of metrics for regression (RMSE, MAE), classification (accuracy, average log-likelihood) and ranking (NDCG@3, NDCG@10). 
For classification, we count a predicted rating to be correct if it is closer to the true rating than to any other value in $\{0.5, 1, \ldots, 5\}$.
For the ranking task, we use models' predicted ratings to rerank each user's test set items with known ratings, taking the average over all users.
Item relevance is \emph{not} binarized as we aim to achieve the best ordering across all ranks.
For regression and ranking, predictions are made using each model's estimate of the rating mean, whilst accuracy is evaluated using the mode of the distribution.
Although different point estimates, e.g., quantiles, may be useful for other tasks, we observed that the above estimates resulted in the highest performance for all models. %

The hyperparameters of each model were determined by an extensive random search. 
All models were trained with embedding sizes: 32, 64, 128, 256 and 512, for up to $50$ epochs with minibatch size of $8192$ using the Adam optimizer \citep{kingma2017Adam} with learning rates in the range $[10^{-6},10^{-2}]$ and early stopping on validation RMSE with a patience of $10$ and a $5\cdot10^{-4}$ tolerance.
We also applied $L_2$ regularization on either all embeddings equally or proportional to their frequency in the training data.
The final parameters per model were chosen according to the best achieved RMSE on the validation set of the first data fold. 
\ac{MF}'s embeddings were used to initialize \ac{CMF} \citep{wang2018ConfidenceAware}.
Furthermore, \citep{koren2011OrdRec} only evaluate a version of their model whose adaptive binning is determined solely on the user (OrdRec-U).
We also include OrdRec with binning based on both the user and item via the exponent from Section~\ref{subsec:binning} (OrdRec-UI).
And while the authors allow for any model to represent the logistic mean, we use OrdRec on top of \ac{MF} to allow for a direct comparison with other confidence models.

To evaluate the effectiveness of the confidence modeling, we measure the correlation between confidence and prediction error using both linear correlation (Pearson's $r$) as well as rank correlation (Kendall's $\tau$).
The former reflects the association between specific error and confidence values, whereas the latter quantifies the effectiveness of using model confidence to separate low and high-error predictions.
Furthermore, we evaluate the relationship qualitatively
by clustering interactions into $1000$ bins based on the predicted variance of individual ratings.
We use rating variance as it is a confidence measure available for all models, and group interactions using equally spaced bins to identify the relationship between accuracy and specific confidence values. 
However, due to this grouping strategy, we may also observe patterns for a small number of interactions spread over many bins that do not generalize to the data as a whole.
Therefore, we also group interactions into bins containing the same number of interactions of monotonically increasing variance, allowing us to evaluate the general direction of the association but not for specific confidence values.

\subsection{High-precision targeted recommendation}
\label{subsec:notification_setup}

Finally, inspired by work on prediction confidence in other domains~\citep{herlocker2000Explaining,mesas2020Exploiting,mazurowski2013Estimating},
we want to evaluate whether the confidence modeling of \ac{LBD} can be effectively exploited in downstream tasks.

We simulate a high-precision task where only a limited number of targeted users are provided with recommendations.
Specifically, each method is only allowed to choose $N$ different users, for each of whom it makes one personalized item recommendation,
while the remaining users receive no recommendations.
For evaluation purposes, the methods may only choose user-item combinations that are present in the test set, so that we can evaluate performance based on the number of successful recommendations, which are taken to be those with rating of at least $4.5$.
For each model, we report the proportion of users who received a successful recommendation (Precision@$1$) over different values of $N$, as well as the relative improvement over the confidence-unaware \ac{MF}.

\begin{table*}[t]
 \caption{
 Performance in terms of regression, classification and ranking metrics for various LBD configurations (see Sections~ \ref{subsec:modeling_mean_confidence}-\ref{subsec:binning}).
Reported results are averages from 10-fold cross-validation, with standard deviations in parentheses.
 Two embedding sizes denote separate embeddings for preference and confidence. 
 $512 + 10$ denotes \ac{LBD} with the adaptive binning strategy applied (LBD-A).
 Highest performance per metric is denoted in bold.
 }
 \label{table:lbd_optimization_results}
  \begin{center}
   \begin{tabular}{cccccccccc}
    \toprule
    \textbf{Size} & $\bm{\nu^{(\cdot)}}$ & \bf Bias &
    \multicolumn{1}{c}{\textbf{RMSE}} & \multicolumn{1}{c}{\textbf{MAE}}  & \multicolumn{1}{c}{\textbf{Accuracy}} & \multicolumn{1}{c}{\textbf{Ave. Log-L.}} & \multicolumn{1}{c}{\textbf{NDCG@3}} & \multicolumn{1}{c}{\textbf{NDCG@10}}\\ \midrule
512 & $sum$ & - &   0.7932 \small{(0.0005)} &   0.6130 \small{(0.0004)} &   0.3016 \small{(0.0005)} &   -1.769 \small{(0.001)} &    0.9330 \small{(0.004)} &    0.9568 \small{(0.0002)} \\
512 & $sum$ & $\mu, \nu$ &  0.7864 \small{(0.0005)} &   0.6058 \small{(0.0003)} &   0.3092 \small{(0.0003)} &   -1.760 \small{(0.001)} &    0.9339 \small{(0.0003)} &   0.9568 \small{(0.0001)} \\
512 & $sum$ & $\alpha, \beta$ & 0.7761 \small{(0.0005)} &   0.5895 \small{(0.0005)} &   0.3139 \small{(0.0003)} &   -1.755 \small{(0.003)} &    \textbf{0.9357} \small{(0.0003)} &  \textbf{0.9583} \small{(0.0001)} \\
512 & $norm$ & $\alpha, \beta$ &    0.7852 \small{(0.0007)} &   0.5942 \small{(0.0005)} &   0.3140 \small{(0.0005)} &   -1.818 \small{(0.004)} &    0.9336 \small{(0.0002)} &   0.9572 \small{(0.0002)} \\
512 & $dot$ & $\alpha, \beta$ & 0.8387 \small{(0.0005)} &   0.6391 \small{(0.0006)} &   0.2972 \small{(0.0005)} &   -1.865 \small{(0.005)} &    0.9330 \small{(0.0004)} &   0.9495 \small{(0.0001)} \\
$256,256$ & $sum$ & $\alpha, \beta$ &   0.7850 \small{(0.0006)} &   0.5964 \small{(0.0006)} &   0.3133 \small{(0.0006)} &   -1.774 \small{(0.002)} &    0.9331 \small{(0.0003)} &   0.9570 \small{(0.0001)} \\
$512 + 10$ & $sum$ & $\alpha, \beta$ &  \textbf{0.7759} \small{(0.0004)} &  \textbf{0.5875} \small{(0.0004)} &  \textbf{0.4356} \small{(0.0006)} &  \textbf{-1.432} \small{(0.002)} &   0.9351 \small{(0.0003)} &   0.9579 \small{(0.0001)} \\
    \bottomrule
   \end{tabular}
  \end{center}
\end{table*}

This task is analogous to a notification setting where recommendations are mildly intrusive interruptions.
For the best user experience in this setting, only highly relevant recommendations should be made and the number of recommendations per user should be limited.
We expect that confidence modeling allows for a higher precision in selecting highly relevant items, since it can better identify which high rating predictions are most likely correct.

\begin{table*}[]
 \caption{
 Model performance on regression, classification and ranking tasks.
 Results are averages from 10-fold cross-validation, with standard deviations in parentheses.
 Highest performance per metric is denoted in bold.
 Significant improvement of an \ac{LBD} model over \emph{all} baseline models is denoted by $\vartriangle$, significant improvement of a baseline over \emph{any} \ac{LBD} version is denoted by $\triangledown$
 (separate one-sided Wilcoxon signed-rank tests with matched folds, $\textit{p}$ < 0.001). %
 }
 \label{table:main_results}
  \begin{center}
   \begin{tabular}{cccccccc}
    \toprule
    \textbf{Model} & \phantom{$^\triangledown$}\textbf{RMSE}\phantom{$^\triangledown$} & \phantom{$^\triangledown$}\textbf{MAE}\phantom{$^\triangledown$}  & \textbf{Accuracy}& \multicolumn{1}{c}{\textbf{Ave. Log-L.}} & \textbf{NDCG@3} & \textbf{NDCG@10}\\ \midrule
MF (128) &  0.7802 \small{(0.0006)} &   0.5984 \small{(0.0005)} &   0.2923 \small{(0.0005)} &   -2.004 \small{(0.002)} &    0.9334 \small{(0.0002)} &   0.9570 \small{(0.0001)} \\
MF (512) &  0.7883 \small{(0.0006)} &   0.6022 \small{(0.0005)} &   0.2934 \small{(0.0004)} &   -2.022 \small{(0.002)} &    0.9321 \small{(0.0002)} &   0.9560 \small{(0.0001)} \\
CMF (128) & 0.7760 \small{(0.0007)} &   0.5936 \small{(0.0004)} &   \phantom{$^\triangledown$}0.3226 \small{(0.0004)} $^\triangledown$ &   -1.777 \small{(0.001)} &    0.9338 \small{(0.0002)} &   0.9573 \small{(0.0001)} \\
CMF (512) & 0.7820 \small{(0.0005)} &   0.5967 \small{(0.0003)} &   0.2849 \small{(0.0015)} &   -1.801 \small{(0.001)} &    0.9329 \small{(0.0003)} &   0.9567 \small{(0.0001)} \\
OrdRec-U (512) & 0.7821 \small{(0.0006)} &   0.6043 \small{(0.0004)} &   0.2322 \small{(0.0007)} &   -1.881 \small{(0.001)} &    0.9343 \small{(0.0003)} &   0.9575 \small{(0.0001)} \\
OrdRec-UI (512) &    0.7765 \small{(0.0006)} &   0.5896 \small{(0.0005)} &   \phantom{$^\triangledown$}0.4187 \small{(0.0006)}$^\triangledown$&  \phantom{$^\triangledown$}-1.569 \small{(0.001)}$^\triangledown$&   0.9349 \small{(0.0003)} &   0.9578 \small{(0.0001)} \\
LBD-S (512) &    0.7761 \small{(0.0005)} &   0.5895 \small{(0.0005)} &   0.3139 \small{(0.0003)} &   -1.755 \small{(0.003)} &    \phantom{$^\triangledown$}\textbf{0.9357} \small{(0.0003)} $^\vartriangle$& \phantom{$^\triangledown$}\textbf{0.9583} \small{(0.0001)}$^\vartriangle$ \\
LBD-A (512) &    \textbf{0.7759} \small{(0.0004)} &  \phantom{$^\triangledown$}\textbf{0.5875} \small{(0.0004)} $^\vartriangle$& \phantom{$^\triangledown$}\textbf{0.4356} \small{(0.0006)} $^\vartriangle$& \phantom{$^\triangledown$}\textbf{-1.432} \small{(0.002)} $^\vartriangle$ & 0.9351 \small{(0.0003)} &   0.9579 \small{(0.0001)} \\
    \bottomrule
   \end{tabular}
  \end{center}
\end{table*} 

\section{Results and Discussion}
\label{sec:results}

\subsection{Parameterization of the \ac{LBD} model}
\label{subsec:rq1}

We begin with addressing the first research question (\ref{rq:parameter}) by investigating \emph{what method of modeling the \ac{LBD} parameters results in the highest predictive performance.}
To that end, we consider Table~\ref{table:lbd_optimization_results} which shows the average performance of various \ac{LBD} configurations over $10$ test folds.
The configurations are combinations of different confidence functions and ($\alpha$, $\beta$) and ($\nu$, $\mu$) biases proposed in Sections~ \ref{subsec:modeling_mean_confidence} and \ref{subsec:biases}.
In addition, we also evaluate a model with no bias terms, a model that uses separate embeddings to model preference and confidence (denoted with $256,256$) and \ac{LBD-A} with adaptively discretized bins (denoted by $512 + 10$, for the ten additional learned binning parameters).

We see that the choice of parameterization can have a substantial impact on performance: overall, using $\nu^\text{sum}$ with $\alpha,\beta$ biases alongside adaptive bins leads to best performance, while using $\nu^\text{dot}$ is clearly the worst choice.
Interestingly, the above model with adaptive bins shows the best performance across all non-ranking metrics, as well as second best performance on ranking, with particularly salient improvements in terms of accuracy and log-likelihood.
Moreover, the same model but without dynamic bins achieves the best performance for ranking metrics as well as the second highest performance on regression metrics and log-likelihood.
This thus suggests that these two parameterizations are best suited to rating prediction, regardless of the specific task.

Based on the observed performance differences, 
it appears that the adaptive binning allows the model to better capture the rating distribution, as evidenced by the associated size of the improvement in the fit to the rating data.
We also hypothesize that $\nu^\text{sum}$ is better at confidence modeling than $\nu^\text{norm}$ due to its ability to capture interactions between specific users and items.
On the other hand, we speculate that the dot product performs the worst due to its high correlation with cosine similarity, which could in turn produce a high correlation between predicted means and corresponding confidences.
Furthermore, our results also indicate that using separate embeddings to model means and confidences does not boost performance, suggesting that a single set of user and item embeddings are enough to capture both preference and confidence patterns.
Lastly, it is also clear that including bias terms can lead to better performance across all metrics, as expected from their popularity for \ac{MF}-based models.
However, we are unsure why bias weights in $\alpha,\beta$ terms are more effective than in $\mu,\nu$ terms.

In conclusion, to answer \ref{rq:parameter}: we find that the parameterization of \ac{LBD} can have a substantial effect on its performance.
\emph{Our results indicate that the confidence function $\nu^\text{sum}$  with $\alpha$ and $\beta$ bias terms, both with and without dynamic binning, lead to the highest overall performance across all our metrics.}

\begin{table*}[t]
 \centering
 \caption{
Correlations between the predicted variance and MAE of baselines and LBD. 
 Results are averages from 10-fold cross-validation, with standard deviations in parentheses.
 Highest correlation per type is denoted in bold.
 Significant improvement of \ac{LBD} over \emph{all} baselines is denoted by $\vartriangle$ (separate one-sided Wilcoxon signed-rank tests with matched folds, $\textit{p}$ < 0.001).
 }
 \label{table:correlation}
   \begin{tabular}{cccccc}
    \toprule
     & CMF & OrdRec-U & OrdRec-UI & LBD-S & LBD-A \\ \midrule
Pearson's $r$ (Linear Correlation) & 0.079 \small{(0.001)} & 0.157 \small{(0.001)} & 0.175 \small{(0.001)} & \phantom{$^\triangledown$}0.304 \small{(0.002)}$^\vartriangle$ & \phantom{$^\triangledown$}\textbf{0.329} \small{(0.002)}$^\vartriangle$\\
Kendall's $\tau$ (Rank Correlation) & 0.043 \small{(0.002)} & 0.094 \small{(0.001)} & 0.112 \small{(0.001)} & \phantom{$^\triangledown$}0.194 \small{(0.001)}$^\vartriangle$ & \phantom{$^\triangledown$}\textbf{0.216} \small{(0.001)}$^\vartriangle$\\
    \bottomrule
   \end{tabular}
\end{table*} 

\begin{figure*}
\setlength{\tabcolsep}{0.04cm}
   \centering
\resizebox{\textwidth}{!}{
   \begin{tabular}{r r r r r r rl}
        \rotatebox{90}{\hspace{5mm}\parbox{2cm}{
            \centering
            Mean Absolute\\
            Error (MAE)
      }}
      & \includegraphics[scale=0.39]{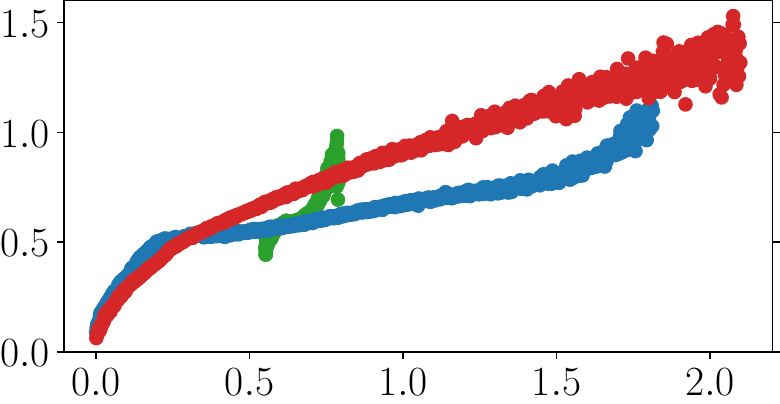}  
      & \includegraphics[scale=0.39]{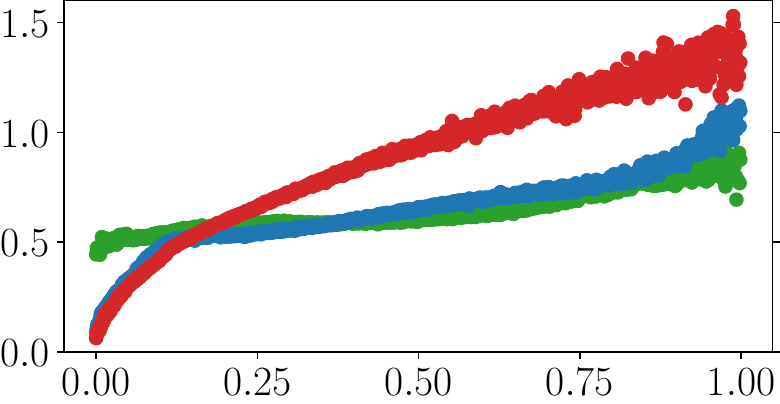}
      & \includegraphics[scale=0.39]{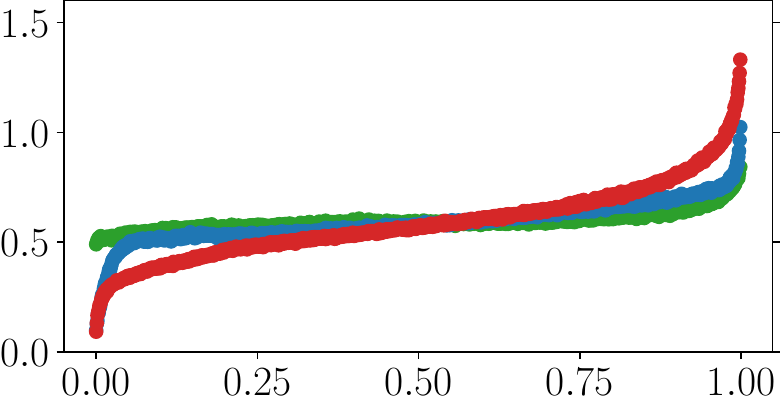}
      & 
      \\
        \rotatebox{90}{\hspace{5mm}\parbox{2cm}{
            \centering
            Mean Predicted\\
            Rating
      }}
        & \includegraphics[scale=0.39]{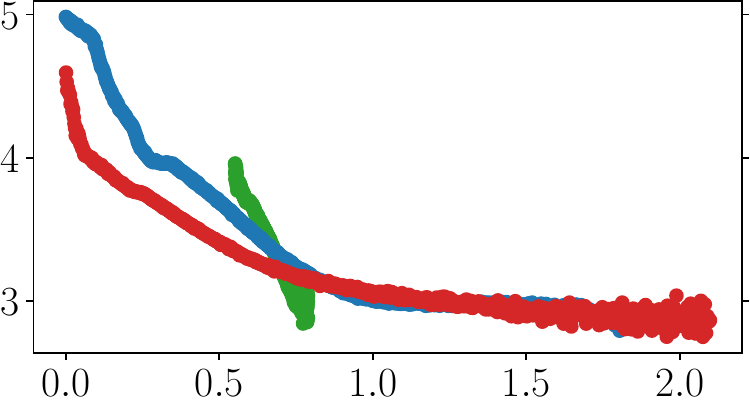}\hspace{0mm}
      &\includegraphics[scale=0.39]{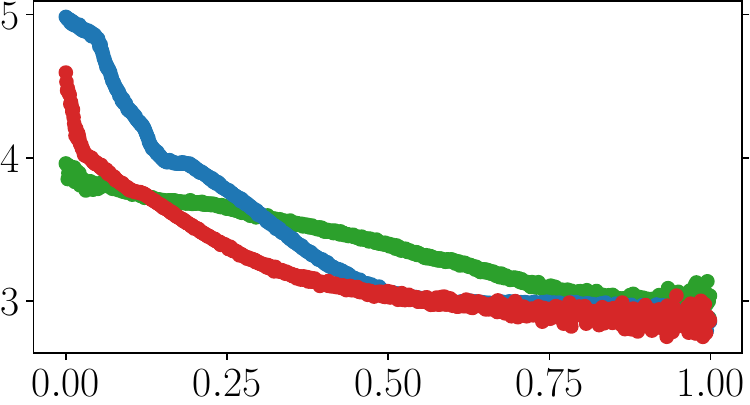}
      & \includegraphics[scale=0.39]{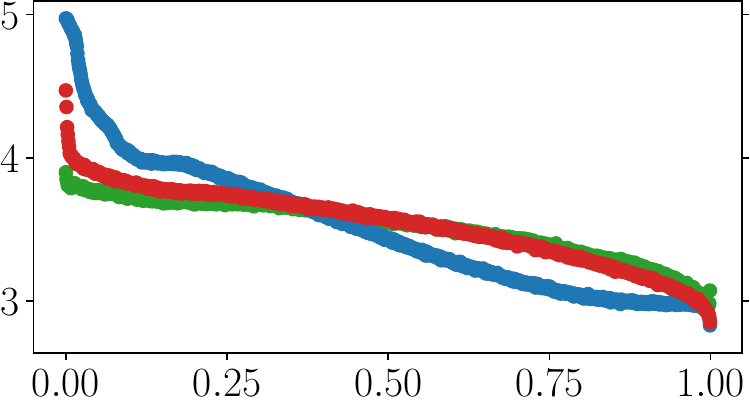}
      &
      \\
      & \multicolumn{1}{c}{\hspace{2mm}Predicted Variance}
      & \multicolumn{1}{c}{\hspace{4.5mm}Predicted Variance Rescaled}
      & \multicolumn{1}{c}{\hspace{5.5mm}Predicted Variance Quantile} &
      \\
      & \multicolumn{3}{c}{\includegraphics[width=0.31\textwidth]{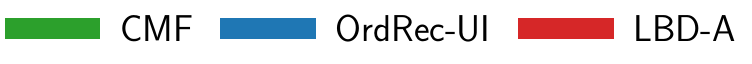}}
   \end{tabular}
   }
   \caption{
        Variation of \ac{MAE} (top) and average predicted rating values (bottom) for different values of predicted variance.
        Interactions grouped by the predicted variance of each model into 1000 clusters, discarding 0.1\% of highest predicted variance outliers.
        Points represent mean response value of each cluster.
        Left: interactions grouped into equispaced bins spanning each model's observed predicted variance. 
        Center: grouping as on the left, but variance rescaled to [0,1].
        Right: each bin contains 0.1\% of interactions, ordered by predicted variance.
      }
   \label{fig:confidence}
\end{figure*}

\subsection{Recommendation performance}
\label{subsec:performance}

Now that we have established the best configurations of the \ac{LBD} model, we can address our second research question (\ref{rq:performance}) and consider \textit{whether the predictive performance of \ac{LBD} is competitive with that of \ac{MF}, \ac{CMF} and OrdRec.}
For this question, we use the results in Table~\ref{table:main_results} which show the performance of all models with different embedding sizes over a variety of performance metrics.

Overall, we generally see no enormous variation in the performance of the models, as most are in a similar range.
Whilst \ac{LBD-S} achieves lower accuracy compared to \ac{CMF} and OrdRec-UI, its regression performance is comparable to that of the above models, and its ranking performance is significantly better than for any other model. 
Furthermore, \ac{LBD-A}, whilst showing a smaller improvement on NDCG, actually achieves the best performance across all other metrics, significantly improving over the baselines in MAE as well as both classification metrics.
Importantly, the above results thus show that \ac{LBD} has performance that is competitive with \ac{MF}, \ac{CMF} and OrdRec.

As such, we answer \ref{rq:performance} in the affirmative: \textit{the performance of \ac{LBD} is competitive with \ac{MF}, \ac{CMF} and OrdRec across all our measured metrics.}
Furthermore, it appears that \ac{LBD} could even provide small significant improvements for most metrics.
We note that our aim is not to improve these performance metrics, but to evaluate whether confidence modeling has a negative effect on predictive performance.
These conclusions thus strongly imply that confidence modeling with \ac{LBD} allows us to simultaneously model both preference and confidence without sacrificing predictive performance, at least not in terms of our regression, classification and ranking metrics.

\subsection{Evaluating prediction confidence}
\label{subsec:rq3}

Now that we have investigated the predictive performance of \ac{LBD}, we turn our attention to its key feature ---its confidence modeling--- and address the third research question (\ref{rq:confidence}) by evaluating \textit{whether there is a stronger correlation between the confidence and accuracy of predictions for \ac{LBD} than for \ac{CMF} and OrdRec}.
We note that no comparison with \ac{MF} can be made, since its pointwise predictions do not include any indication of confidence, uncertainty or variance. 
Table~\ref{table:correlation} shows the strength of linear and rank correlation between \ac{MAE} and predicted variance,
representing the association between the prediction errors and confidence.
We use predicted variance to quantify confidence, since low variance denotes a prediction with low uncertainty and thus high confidence, and vice versa.

From Table~\ref{table:correlation}, it is immediately clear that there are substantial differences across models in terms of both types of correlation.
Whilst OrdRec-UI's correlation is the highest among the baselines, \ac{LBD-S} and \ac{LBD-A} demonstrate by far the strongest association between their predicted variance and the size of the error.
In fact, compared to OrdRec-UI, both \ac{LBD} models demonstrate a relative increase in both correlation coefficients of at least $70\%$, with \ac{LBD-A}'s correlation being almost twice as strong as that of OrdRec.
That \ac{LBD-S} achives such a large improvement over OrdRec-UI indicates that modeling user preferences with just a beta distribution and no adaptive binning better captures uncertainty in user ratings compared to the non-linear mapping of OrdRec.
Overall, the above results thus suggest that \ac{LBD} is considerably better at separating low-error predictions from those with a higher error.

For a more qualitative understanding of the above differences, Figure~\ref{fig:confidence} shows how \ac{MAE} and the average predicted rating vary over interactions grouped by predicted variance.
Furthermore, as the predicted variance of distinct models lies over different ranges (see left column of Figure~\ref{fig:confidence}),
we also show the correlation with predicted variance rescaled to the range $[0,1]$ (center column), and variance quantiles (right column).
We omit OrdRec-U and \ac{LBD-S} from the visualization to reduce visual clutter.

\begin{table*}[t]
 \centering
 \caption{
 Model performance on the high-precision targeted recommendation task for distinct numbers of targeted users ($N$, half-powers of 10).
 Results are Precision@1 averages from 10-fold cross-validation, with standard deviations in parentheses.
 Highest performance for each value of $N$ is denoted in bold.
 Significant improvement of an \ac{LBD} model over \emph{all} baseline models is denoted by $\vartriangle$, no significant improvement of any of the baselines over either \ac{LBD} model  (separate one-sided Wilcoxon signed-rank tests with matched folds, \textit{p <} 0.001).
 }
 \label{table:notification}
   \begin{tabular}{ccccccc}
    \toprule
    \textbf{Model}& \multicolumn{6}{c}{\textbf{Precision@$1$ for top $N$ users}} \\
     & $N=100$ & $N=320$ & $N=1000$ & $N=3200$ & $N=10000$ & $N=32000 $\\ \midrule
MF &  0.885 \small{(0.034)} &  0.893 \small{(0.019)} &  0.889 \small{(0.011)} &  0.886 \small{(0.004)} &  0.841 \small{(0.004)} &  0.726 \small{(0.002)}\\
CMF & 0.913 \small{(0.030)} &  0.907 \small{(0.016)} &  0.904 \small{(0.013)} &  0.895 \small{(0.005)} &  0.853 \small{(0.005)} &  0.733 \small{(0.002)}\\
OrdRec-UI &   0.962 \small{(0.016)} &  0.949 \small{(0.007)} &  0.932 \small{(0.006)} &  0.910 \small{(0.003)} &  0.863 \small{(0.003)} &  0.747 \small{(0.002)}\\
LBD-S &   0.947 \small{(0.019)} &  0.944 \small{(0.013)} &  0.932 \small{(0.009)} &  0.908 \small{(0.006)} &  0.861 \small{(0.004)} &  0.747 \small{(0.003)}\\
LBD-A &   \textbf{0.971} \small{(0.012)} &  \phantom{$^\vartriangle$}\textbf{0.962} \small{(0.011)}$^\vartriangle$ &  \phantom{$^\vartriangle$}\textbf{0.949} \small{(0.005)}$^\vartriangle$ &  \phantom{$^\vartriangle$}\textbf{0.925} \small{(0.004)}$^\vartriangle$ &  \phantom{$^\vartriangle$}\textbf{0.877} \small{(0.004)}$^\vartriangle$ &  \phantom{$^\vartriangle$}\textbf{0.751} \small{(0.003)}$^\vartriangle$ \\
    \bottomrule
   \end{tabular}
\end{table*} 

\begin{figure*}
   \centering
   \begin{tabular}{r l r l}
      \rotatebox{90}{
         \hspace{0.72cm}
         \parbox{2cm}{
            \centering
            Precision@$1$
      }}
      &
      \includegraphics[scale=0.485]{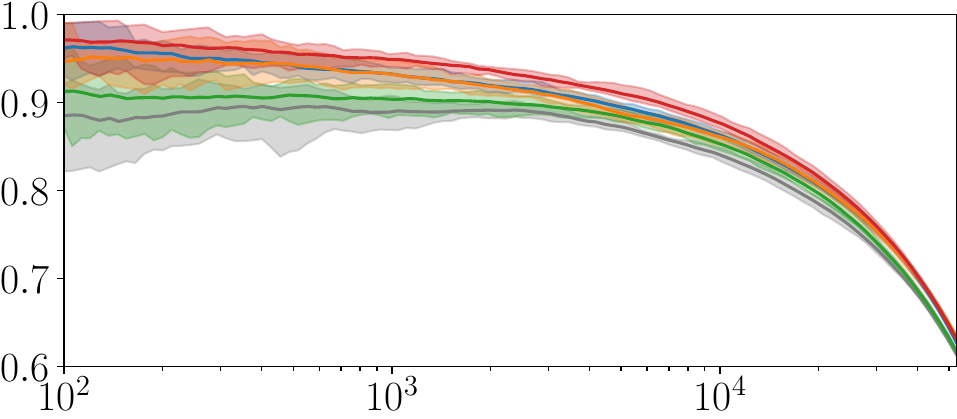}
      &
      \rotatebox{90}{
         \hspace{0.72cm}
         \parbox{2cm}{
            \centering
            \% Gain vs MF
      }}
      &
      \includegraphics[scale=0.485]{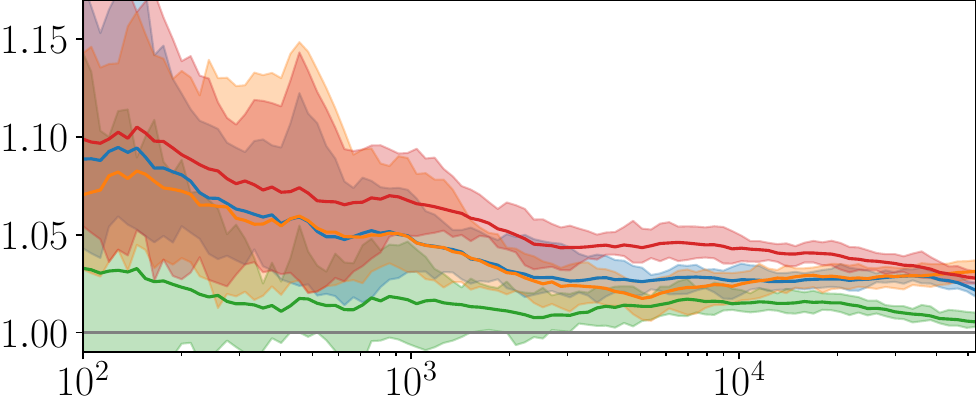}
      \\
      & \multicolumn{1}{c}{\# Users with a Recommendation ($N$)} & & \multicolumn{1}{c}{\# Users with a Recommendation ($N$)}
      \\
     \multicolumn{4}{c}{\includegraphics[width=0.523\textwidth]{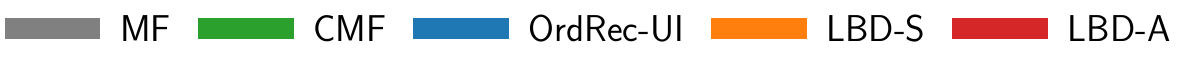}}
   \end{tabular}
   \caption{
      Model performance on the high-precision targeted recommendation task for distinct numbers of targeted users ($N$) with \ac{MF}, \ac{CMF}, OrdRec and \ac{LBD} methods.
      Left: Percentage of users for whom a 4.5+ star recommendation is made (Precision@1).
      Right: Relative Precision@1 improvement over Matrix Factorization.
      Results are averages over 10 folds. Shaded regions denote minimum and maximum across folds.
      }%
   \label{fig:notification_results}
\end{figure*}

We start by considering the top-left of Figure~\ref{fig:confidence}, which shows how the predicted variance of \ac{CMF}, \ac{LBD-A} and OrdRec-UI correlates with their accuracy, with higher variance corresponding to greater \ac{MAE} for all.
Rescaling each model's variance for a more direct comparison, as done in the top-center of Figure~\ref{fig:confidence}, shows that there is a larger difference in terms of \ac{MAE} between the lowest and highest variance predictions of \ac{LBD-A} and OrdRec-UI compared to \ac{CMF}.
This suggests that \ac{LBD-A} and OrdRec-UI are more efficient at separating high and low-error predictions.

Interestingly, from top-center of Figure~\ref{fig:confidence}, one might also (albeit erroneously) conclude that, 
because \ac{MAE} of OrdRec-UI and \ac{LBD-A} is comparable when their predicted variance is low and otherwise generally lower for OrdRec-UI, that must mean that OrdRec-UI's confidence modeling is better.
We can see that this is however not true by inspecting the top-right of Figure~\ref{fig:confidence}, which presents the same results but with predictions ordered in quantiles of variance.
There we see that using the predicted variance of \ac{LBD-A} achieves a better separation between high and low \ac{MAE} predictions when compared to OrdRec-UI (and \ac{CMF}).
OrdRec does appear to achieve a similar error to \ac{LBD-A} on the interactions with the very lowest predicted variance, whilst also having the highest error in the right tail.
However, the majority of the interactions, which are in the center, all have only minor differences in their \ac{MAE}.
\ac{LBD-A}, on the other hand, whilst being particularly efficient in the tails, is also better at separating all of the other interactions.
As such, for a comparable level of predictive performance (Section~\ref{subsec:performance}), \ac{LBD} is better at predicting when it is more or less accurate, i.e. it has a substantially better model of confidence.

To get further insight into the different confidence models, the bottom row of Figure~\ref{fig:confidence} also displays how the average predicted rating varies over the predicted variances.
We see that all three models have more confidence in high value ratings, with this trend being the strongest for OrdRec.
Potentially, this is an indication that it is easier to predict the positive preferences of a user than their neutral or negative preference.
However, for the \ac{LBD} model, we speculate that the usage of the $\nu^\text{sum}$ confidence function could also produce this trend.
It assigns higher confidence when embeddings have a similar direction, and similar directions also result in higher ratings due to the usage of cosine similarity.
Due to the scope of this work, we have to leave a further investigation into the implications of these trends for future work.

Altogether, we answer \ref{rq:confidence} strongly in the affirmative: \emph{\ac{LBD} provides a significantly and considerably stronger correlation between the confidence and accuracy of predictions than \ac{CMF} or OrdRec.}
Consequently, \ac{LBD} is substantially better at separating accurate predictions from those with moderate or high prediction errors.

\subsection{High-precision targeted recommendation}
\label{subsec:notification}

Finally, in order to investigate whether the better confidence modeling of \ac{LBD} could translate into meaningful improvements for downstream tasks, we turn to \ref{rq:downstream} and ask \emph{whether LBD can reach higher performance in a high-precision recommendation task}.

For this question, we simulate a notification task where each user can only be recommended a single item, and due to the mild intrusiveness of a notification, only $N$ total users will receive a recommendation.
Therefore, this task could be divided into two stages: first to recognize what is the best item to recommend for each user, and second, to determine which $N$ recommendations of the first stage are most likely to result in a positive reaction.
We expect that an accurate confidence model is better suited for both stages of this task.

We divide our simulation in these stages: first, for each user we select the test set item that has the highest predicted probability of a 4.5+ star rating (i.e., for models not relying on binning: the probability of being at least $4.25$).
For \ac{MF}, this is also equivalent to simply using its predicted rating value~\citep{mnih2007Probabilistic}.
We then use the above estimates to order the resulting user-item combinations. 
Then, for a given $N$, we calculate what percentage of the top-$N$ user-item combinations truly did receive a 4.5+ star rating in the test set (Precision@1), i.e., the number of users who are happy with their notification. 
Note that $N$ does not exceed the true number of users who have a 4.5+ star test set rating.

Figure~\ref{fig:notification_results} shows the results for this experiment in terms of Precision@1 for various values of $N$ as well as models' relative improvement in the metric over \ac{MF}.
From the figure we see that all methods can identify a highly relevant item for most users, especially when $N$ is small, with the task becoming progressively more difficult as $N$ increases.
Surprisingly, our results only show minor differences between \ac{CMF} and \ac{MF}, with the relative improvement of \ac{CMF} limited to below $2\%$ for most values of $N$.
In contrast, OrdRec-UI, \ac{LBD-S} and \ac{LBD-A} start off with a substantially higher precision@$1$ for $N=100$ and consistently outperform \ac{CMF} even as $N$ increases and users with lower confidence are included.
Model performance for a selection of user numbers is also shown in Table~\ref{table:notification}. 
We can see that the performance of \ac{LBD-S} is similar to that of OrdRec-UI, with frequently overlapping means and no significant improvement of one model over other.
It appears that OrdRec-UI's weaker confidence modeling may still be quite efficient at identifying low-error high-rating items.
\ac{LBD-A}, on the other hand, consistently achieves the highest performance, with particularly noticeable improvement over OrdRec-UI of approximately $1.5\%$ (absolute differences) for $1000\leq N \leq 10{,}000$.
As the improvements of \ac{LBD-A} are also significant for all listed values $N>100$, we conclude that \ac{LBD-A} is meaningfully better at this high-precision targeted recommendation task, likely due to its better confidence modeling capabilities.

As such, we also answer \textbf{RQ4} in the affirmative: \emph{\ac{LBD}, and particularly \ac{LBD-A}, brings a meaningful performance improvement for the high-precision targeted recommendation task over \ac{MF}, \ac{CMF} and OrdRec}.
Combined with the conclusion from \textbf{RQ3}, our results strongly indicate that the confidence model of \ac{LBD} is both a better indication of its prediction accuracy and that this higher accuracy can be translated to meaningful benefits for downstream tasks.

\section{Conclusion and Future Work}

In this work, we investigated whether confidence modeling for \ac{RecSys} is possible while maintaining high predictive accuracy and without introducing high model-complexity or heavy computational costs.
To this end, we proposed \acf{LBD} as a simple \ac{RecSys} with an explicit model of confidence while having the same model-complexity as \ac{MF}.

Our experimental results show that \ac{LBD} has competitive predictive performance with \ac{MF}, \ac{CMF} and OrdRec, where no decreases were observed for any metric.
Furthermore, unlike \ac{CMF} and OrdRec, we found that the confidence of \ac{LBD} correlates strongly with its accuracy, allowing it to better separate predictions on their accuracy.
Finally, we found that \ac{LBD}'s superior  confidence modeling translates to a meaningfully higher precision in a high-precision targeted recommendation task, compared to the other methods.

We conclude that with the introduction of \ac{LBD}, we have shown that accurate recommendation can be combined with accurate confidence modeling and low model-complexity.
Furthermore, our findings indicate that such confidence models can lead to meaningful improvements in downstream tasks: \ac{LBD} thereby provides a very promising potential for novel future applications.

Future work could investigate whether the confidence model of \ac{LBD} can be used to better understand user preferences, or the uncertainty contained in datasets.
While this work focused on matching the model-complexity of \ac{MF}, the concept of \ac{LBD} is easily applicable or extendable to more complex \ac{RecSys} models.
Similarly, while we tackled rating-prediction, \ac{LBD} could be extended to confidence modeling for predictions of implicit feedback signals, e.g., clicks, purchases or watchtime.
Additionally, it would be of interest to better understand how confidence modeling of LBD interacts with various cognitive and statistical biases, and the approaches used to correct them \citep{knyazev2022bandwagon,Schnabel2016,huang2022It,swaminathan2015Self}.
Finally,  \ac{LBD} is designed to be practical for real-world settings; therefore, we hope practitioners will consider \ac{LBD} and its many potential applications.

\begin{acks}
This work used the Dutch national e-infrastructure with the support of the SURF Cooperative using grant no. EINF-4538.
\end{acks}

\subsection*{Code and data}
To facilitate the reproducibility of the reported results, this work only made use of publicly available data and our experimental implementation is publicly available at \url{https://github.com/NKNY/confidencerecsys2023}.

\balance
\bibliographystyle{ACM-Reference-Format}
\bibliography{confidencerecsys2023}


\begin{thebibliography}{52}


\ifx \showCODEN    \undefined \def \showCODEN     #1{\unskip}     \fi
\ifx \showDOI      \undefined \def \showDOI       #1{#1}\fi
\ifx \showISBNx    \undefined \def \showISBNx     #1{\unskip}     \fi
\ifx \showISBNxiii \undefined \def \showISBNxiii  #1{\unskip}     \fi
\ifx \showISSN     \undefined \def \showISSN      #1{\unskip}     \fi
\ifx \showLCCN     \undefined \def \showLCCN      #1{\unskip}     \fi
\ifx \shownote     \undefined \def \shownote      #1{#1}          \fi
\ifx \showarticletitle \undefined \def \showarticletitle #1{#1}   \fi
\ifx \showURL      \undefined \def \showURL       {\relax}        \fi
\providecommand\bibfield[2]{#2}
\providecommand\bibinfo[2]{#2}
\providecommand\natexlab[1]{#1}
\providecommand\showeprint[2][]{arXiv:#2}

\bibitem[Abadi et~al\mbox{.}(2015)]%
        {tensorflow2015}
\bibfield{author}{\bibinfo{person}{Mart\'{i}n Abadi}, \bibinfo{person}{Ashish
  Agarwal}, \bibinfo{person}{Paul Barham}, \bibinfo{person}{Eugene Brevdo},
  \bibinfo{person}{Zhifeng Chen}, \bibinfo{person}{Craig Citro},
  \bibinfo{person}{Greg~S. Corrado}, \bibinfo{person}{Andy Davis},
  \bibinfo{person}{Jeffrey Dean}, \bibinfo{person}{Matthieu Devin},
  \bibinfo{person}{Sanjay Ghemawat}, \bibinfo{person}{Ian Goodfellow},
  \bibinfo{person}{Andrew Harp}, \bibinfo{person}{Geoffrey Irving},
  \bibinfo{person}{Michael Isard}, \bibinfo{person}{Yangqing Jia},
  \bibinfo{person}{Rafal Jozefowicz}, \bibinfo{person}{Lukasz Kaiser},
  \bibinfo{person}{Manjunath Kudlur}, \bibinfo{person}{Josh Levenberg},
  \bibinfo{person}{Dandelion Man\'{e}}, \bibinfo{person}{Rajat Monga},
  \bibinfo{person}{Sherry Moore}, \bibinfo{person}{Derek Murray},
  \bibinfo{person}{Chris Olah}, \bibinfo{person}{Mike Schuster},
  \bibinfo{person}{Jonathon Shlens}, \bibinfo{person}{Benoit Steiner},
  \bibinfo{person}{Ilya Sutskever}, \bibinfo{person}{Kunal Talwar},
  \bibinfo{person}{Paul Tucker}, \bibinfo{person}{Vincent Vanhoucke},
  \bibinfo{person}{Vijay Vasudevan}, \bibinfo{person}{Fernanda Vi\'{e}gas},
  \bibinfo{person}{Oriol Vinyals}, \bibinfo{person}{Pete Warden},
  \bibinfo{person}{Martin Wattenberg}, \bibinfo{person}{Martin Wicke},
  \bibinfo{person}{Yuan Yu}, {and} \bibinfo{person}{Xiaoqiang Zheng}.}
  \bibinfo{year}{2015}\natexlab{}.
\newblock \bibinfo{title}{{TensorFlow}: Large-Scale Machine Learning on
  Heterogeneous Systems}.
\newblock
\newblock


\bibitem[Adomavicius et~al\mbox{.}(2007)]%
        {adomavicius2007More}
\bibfield{author}{\bibinfo{person}{Gediminas Adomavicius},
  \bibinfo{person}{Sreeharsha Kamireddy}, {and} \bibinfo{person}{YoungOk
  Kwon}.} \bibinfo{year}{2007}\natexlab{}.
\newblock \showarticletitle{Towards {{More Confident Recommendations}}:
  {{Improving Recommender Systems Using Filtering Approach Based}} on {{Rating
  Variance}}}.
\newblock  (\bibinfo{year}{2007}), \bibinfo{pages}{6}.
\newblock


\bibitem[Anelli et~al\mbox{.}(2020)]%
        {anelli2020recsys}
\bibfield{author}{\bibinfo{person}{Vito~Walter Anelli}, \bibinfo{person}{Amra
  Deli{\'c}}, \bibinfo{person}{Gabriele Sottocornola}, \bibinfo{person}{Jessie
  Smith}, \bibinfo{person}{Nazareno Andrade}, \bibinfo{person}{Luca Belli},
  \bibinfo{person}{Michael Bronstein}, \bibinfo{person}{Akshay Gupta},
  \bibinfo{person}{Sofia Ira~Ktena}, \bibinfo{person}{Alexandre Lung-Yut-Fong},
  {et~al\mbox{.}}} \bibinfo{year}{2020}\natexlab{}.
\newblock \showarticletitle{RecSys 2020 challenge workshop: engagement
  prediction on Twitter’s home timeline}. In
  \bibinfo{booktitle}{\emph{Fourteenth ACM Conference on Recommender Systems}}.
  \bibinfo{pages}{623--627}.
\newblock


\bibitem[Bernardis et~al\mbox{.}(2019)]%
        {bernardis2019Estimating}
\bibfield{author}{\bibinfo{person}{Cesare Bernardis}, \bibinfo{person}{Maurizio
  Ferrari~Dacrema}, {and} \bibinfo{person}{Paolo Cremonesi}.}
  \bibinfo{year}{2019}\natexlab{}.
\newblock \showarticletitle{Estimating {{Confidence}} of {{Individual User
  Predictions}} in {{Item-based Recommender Systems}}}. In
  \bibinfo{booktitle}{\emph{Proceedings of the 27th {{ACM Conference}} on
  {{User Modeling}}, {{Adaptation}} and {{Personalization}}}}.
  \bibinfo{publisher}{{Association for Computing Machinery}},
  \bibinfo{pages}{149--156}.
\newblock


\bibitem[Bouneffouf et~al\mbox{.}(2013)]%
        {bouneffouf2013RiskAwarea}
\bibfield{author}{\bibinfo{person}{Djallel Bouneffouf}, \bibinfo{person}{Amel
  Bouzeghoub}, {and} \bibinfo{person}{Alda~Lopes Ganarski}.}
  \bibinfo{year}{2013}\natexlab{}.
\newblock \showarticletitle{Risk-{{Aware Recommender Systems}}}. In
  \bibinfo{booktitle}{\emph{Neural {{Information Processing}}}},
  \bibfield{editor}{\bibinfo{person}{Minho Lee}, \bibinfo{person}{Akira
  Hirose}, \bibinfo{person}{Zeng-Guang Hou}, {and} \bibinfo{person}{Rhee~Man
  Kil}} (Eds.). \bibinfo{publisher}{{Springer}}, \bibinfo{pages}{57--65}.
\newblock


\bibitem[Bradbury et~al\mbox{.}(2018)]%
        {jax2018}
\bibfield{author}{\bibinfo{person}{James Bradbury}, \bibinfo{person}{Roy
  Frostig}, \bibinfo{person}{Peter Hawkins}, \bibinfo{person}{Matthew~James
  Johnson}, \bibinfo{person}{Chris Leary}, \bibinfo{person}{Dougal Maclaurin},
  \bibinfo{person}{George Necula}, \bibinfo{person}{Adam Paszke},
  \bibinfo{person}{Jake Vander{P}las}, \bibinfo{person}{Skye
  Wanderman-{M}ilne}, {and} \bibinfo{person}{Qiao Zhang}.}
  \bibinfo{year}{2018}\natexlab{}.
\newblock \bibinfo{booktitle}{\emph{{JAX}: composable transformations of
  {P}ython+{N}um{P}y programs}}.
\newblock


\bibitem[Casella and George(1992)]%
        {casella1992Explaining}
\bibfield{author}{\bibinfo{person}{George Casella} {and}
  \bibinfo{person}{Edward~I. George}.} \bibinfo{year}{1992}\natexlab{}.
\newblock \showarticletitle{Explaining the {{Gibbs Sampler}}}.
\newblock \bibinfo{journal}{\emph{The American Statistician}}
  \bibinfo{volume}{46}, \bibinfo{number}{3} (\bibinfo{year}{1992}),
  \bibinfo{pages}{167--174}.
\newblock
\showISSN{0003-1305}


\bibitem[Chu et~al\mbox{.}(2011)]%
        {chu2011Contextual}
\bibfield{author}{\bibinfo{person}{Wei Chu}, \bibinfo{person}{Lihong Li},
  \bibinfo{person}{Lev Reyzin}, {and} \bibinfo{person}{Robert Schapire}.}
  \bibinfo{year}{2011}\natexlab{}.
\newblock \showarticletitle{Contextual {{Bandits}} with {{Linear Payoff
  Functions}}}. In \bibinfo{booktitle}{\emph{Proceedings of the {{Fourteenth
  International Conference}} on {{Artificial Intelligence}} and
  {{Statistics}}}}. \bibinfo{publisher}{{JMLR Workshop and Conference
  Proceedings}}, \bibinfo{pages}{208--214}.
\newblock
\showISSN{1938-7228}


\bibitem[Cuyt et~al\mbox{.}(2008)]%
        {cuyt2008Handbook}
\bibfield{editor}{\bibinfo{person}{Annie Cuyt}, \bibinfo{person}{Vigdis~Brevik
  Petersen}, \bibinfo{person}{Brigitte Verdonk}, \bibinfo{person}{Haakon
  Waadeland}, {and} \bibinfo{person}{William~B. Jones}} (Eds.).
  \bibinfo{year}{2008}\natexlab{}.
\newblock \bibinfo{booktitle}{\emph{Handbook of Continued Fractions for Special
  Functions}}.
\newblock \bibinfo{publisher}{{Springer}}.
\newblock


\bibitem[Gupta and Nadarajah(2004)]%
        {gupta2004handbook}
\bibfield{author}{\bibinfo{person}{Arjun~K Gupta} {and}
  \bibinfo{person}{Saralees Nadarajah}.} \bibinfo{year}{2004}\natexlab{}.
\newblock \bibinfo{booktitle}{\emph{Handbook of beta distribution and its
  applications}}.
\newblock \bibinfo{publisher}{CRC press}.
\newblock


\bibitem[Harper and Konstan(2015)]%
        {harper2015movielens}
\bibfield{author}{\bibinfo{person}{F~Maxwell Harper} {and}
  \bibinfo{person}{Joseph~A Konstan}.} \bibinfo{year}{2015}\natexlab{}.
\newblock \showarticletitle{The movielens datasets: History and context}.
\newblock \bibinfo{journal}{\emph{Acm transactions on interactive intelligent
  systems (tiis)}} \bibinfo{volume}{5}, \bibinfo{number}{4}
  (\bibinfo{year}{2015}), \bibinfo{pages}{1--19}.
\newblock


\bibitem[Herlocker et~al\mbox{.}(2000)]%
        {herlocker2000Explaining}
\bibfield{author}{\bibinfo{person}{Jonathan~L. Herlocker},
  \bibinfo{person}{Joseph~A. Konstan}, {and} \bibinfo{person}{John Riedl}.}
  \bibinfo{year}{2000}\natexlab{}.
\newblock \showarticletitle{Explaining Collaborative Filtering
  Recommendations}. In \bibinfo{booktitle}{\emph{Proceedings of the 2000
  {{ACM}} Conference on {{Computer}} Supported Cooperative Work}}.
  \bibinfo{publisher}{{ACM}}, \bibinfo{pages}{241--250}.
\newblock


\bibitem[Huang et~al\mbox{.}(2022)]%
        {huang2022It}
\bibfield{author}{\bibinfo{person}{Jin Huang}, \bibinfo{person}{Harrie
  Oosterhuis}, {and} \bibinfo{person}{Maarten De~Rijke}.}
  \bibinfo{year}{2022}\natexlab{}.
\newblock \showarticletitle{It {{Is Different When Items Are Older}}:
  {{Debiasing Recommendations When Selection Bias}} and {{User Preferences Are
  Dynamic}}}. In \bibinfo{booktitle}{\emph{Proceedings of the {{Fifteenth ACM
  International Conference}} on {{Web Search}} and {{Data Mining}}}}.
  \bibinfo{publisher}{{ACM}}, \bibinfo{pages}{381--389}.
\newblock


\bibitem[H{\"u}llermeier and Waegeman(2021)]%
        {hullermeier2021aleatoric}
\bibfield{author}{\bibinfo{person}{Eyke H{\"u}llermeier} {and}
  \bibinfo{person}{Willem Waegeman}.} \bibinfo{year}{2021}\natexlab{}.
\newblock \showarticletitle{Aleatoric and epistemic uncertainty in machine
  learning: An introduction to concepts and methods}.
\newblock \bibinfo{journal}{\emph{Machine Learning}} \bibinfo{volume}{110},
  \bibinfo{number}{3} (\bibinfo{year}{2021}), \bibinfo{pages}{457--506}.
\newblock


\bibitem[Jeunen and Goethals(2021)]%
        {jeunen2021Pessimistic}
\bibfield{author}{\bibinfo{person}{Olivier Jeunen} {and} \bibinfo{person}{Bart
  Goethals}.} \bibinfo{year}{2021}\natexlab{}.
\newblock \showarticletitle{Pessimistic {{Reward Models}} for {{Off-Policy
  Learning}} in {{Recommendation}}}.
\newblock In \bibinfo{booktitle}{\emph{Fifteenth {{ACM Conference}} on
  {{Recommender Systems}}}}. \bibinfo{pages}{63--74}.
\newblock


\bibitem[Jin et~al\mbox{.}(2003)]%
        {jin2003Collaborative}
\bibfield{author}{\bibinfo{person}{Rong Jin}, \bibinfo{person}{Luo Si},
  \bibinfo{person}{ChengXiang Zhai}, {and} \bibinfo{person}{Jamie Callan}.}
  \bibinfo{year}{2003}\natexlab{}.
\newblock \showarticletitle{Collaborative Filtering with Decoupled Models for
  Preferences and Ratings}. In \bibinfo{booktitle}{\emph{Proceedings of the
  Twelfth International Conference on {{Information}} and Knowledge
  Management}}. \bibinfo{publisher}{{ACM}}, \bibinfo{pages}{309--316}.
\newblock


\bibitem[Johnson(2014)]%
        {johnson2014logistic}
\bibfield{author}{\bibinfo{person}{Christopher~C Johnson}.}
  \bibinfo{year}{2014}\natexlab{}.
\newblock \showarticletitle{Logistic matrix factorization for implicit feedback
  data}.
\newblock \bibinfo{journal}{\emph{Advances in Neural Information Processing
  Systems}} \bibinfo{volume}{27}, \bibinfo{number}{78} (\bibinfo{year}{2014}),
  \bibinfo{pages}{1--9}.
\newblock


\bibitem[Johnson et~al\mbox{.}(1994)]%
        {johnson1994beta}
\bibfield{author}{\bibinfo{person}{Norman~L Johnson}, \bibinfo{person}{Samuel
  Kotz}, {and} \bibinfo{person}{N Balakrishnan}.}
  \bibinfo{year}{1994}\natexlab{}.
\newblock \showarticletitle{Beta distributions}.
\newblock \bibinfo{journal}{\emph{Continuous univariate distributions. 2nd ed.
  New York, NY: John Wiley and Sons}} (\bibinfo{year}{1994}),
  \bibinfo{pages}{221--235}.
\newblock


\bibitem[Khan et~al\mbox{.}(2021)]%
        {khan2021Deep}
\bibfield{author}{\bibinfo{person}{Zahid~Younas Khan},
  \bibinfo{person}{Zhendong Niu}, \bibinfo{person}{Sulis Sandiwarno}, {and}
  \bibinfo{person}{Rukundo Prince}.} \bibinfo{year}{2021}\natexlab{}.
\newblock \showarticletitle{Deep Learning Techniques for Rating Prediction: A
  Survey of the State-of-the-Art}.
\newblock \bibinfo{journal}{\emph{Artificial Intelligence Review}}
  \bibinfo{volume}{54}, \bibinfo{number}{1} (\bibinfo{year}{2021}),
  \bibinfo{pages}{95--135}.
\newblock
\showISSN{1573-7462}


\bibitem[Kingma and Ba(2017)]%
        {kingma2017Adam}
\bibfield{author}{\bibinfo{person}{Diederik~P. Kingma} {and}
  \bibinfo{person}{Jimmy Ba}.} \bibinfo{year}{2017}\natexlab{}.
\newblock \bibinfo{title}{Adam: {{A Method}} for {{Stochastic Optimization}}}.
\newblock
\newblock
\showeprint[arxiv]{1412.6980}~[cs]


\bibitem[Kl{\"a}s and Vollmer(2018)]%
        {klas2018uncertainty}
\bibfield{author}{\bibinfo{person}{Michael Kl{\"a}s} {and}
  \bibinfo{person}{Anna~Maria Vollmer}.} \bibinfo{year}{2018}\natexlab{}.
\newblock \showarticletitle{Uncertainty in machine learning applications: A
  practice-driven classification of uncertainty}. In
  \bibinfo{booktitle}{\emph{International conference on computer safety,
  reliability, and security}}. Springer, \bibinfo{pages}{431--438}.
\newblock


\bibitem[Knijnenburg et~al\mbox{.}(2012)]%
        {knijnenburg2012explaining}
\bibfield{author}{\bibinfo{person}{Bart~P Knijnenburg},
  \bibinfo{person}{Martijn~C Willemsen}, \bibinfo{person}{Zeno Gantner},
  \bibinfo{person}{Hakan Soncu}, {and} \bibinfo{person}{Chris Newell}.}
  \bibinfo{year}{2012}\natexlab{}.
\newblock \showarticletitle{Explaining the user experience of recommender
  systems}.
\newblock \bibinfo{journal}{\emph{User modeling and user-adapted interaction}}
  \bibinfo{volume}{22}, \bibinfo{number}{4} (\bibinfo{year}{2012}),
  \bibinfo{pages}{441--504}.
\newblock


\bibitem[Knyazev and Oosterhuis(2022)]%
        {knyazev2022bandwagon}
\bibfield{author}{\bibinfo{person}{Norman Knyazev} {and}
  \bibinfo{person}{Harrie Oosterhuis}.} \bibinfo{year}{2022}\natexlab{}.
\newblock \showarticletitle{The Bandwagon Effect: Not Just Another Bias}. In
  \bibinfo{booktitle}{\emph{Proceedings of the 2022 ACM SIGIR International
  Conference on the Theory of Information Retrieval}}.
\newblock


\bibitem[Koren(2009)]%
        {koren2009Collaborative}
\bibfield{author}{\bibinfo{person}{Yehuda Koren}.}
  \bibinfo{year}{2009}\natexlab{}.
\newblock \showarticletitle{Collaborative Filtering with Temporal Dynamics}. In
  \bibinfo{booktitle}{\emph{Proceedings of the 15th {{ACM SIGKDD}}
  International Conference on {{Knowledge}} Discovery and Data Mining}}.
  \bibinfo{publisher}{{ACM}}, \bibinfo{pages}{447--456}.
\newblock


\bibitem[Koren and Bell(2011)]%
        {koren2011Advances}
\bibfield{author}{\bibinfo{person}{Yehuda Koren} {and} \bibinfo{person}{Robert
  Bell}.} \bibinfo{year}{2011}\natexlab{}.
\newblock \showarticletitle{Advances in {{Collaborative Filtering}}}.
\newblock In \bibinfo{booktitle}{\emph{Recommender {{Systems Handbook}}}},
  \bibfield{editor}{\bibinfo{person}{Francesco Ricci}, \bibinfo{person}{Lior
  Rokach}, \bibinfo{person}{Bracha Shapira}, {and} \bibinfo{person}{Paul~B.
  Kantor}} (Eds.). \bibinfo{publisher}{{Springer US}},
  \bibinfo{pages}{145--186}.
\newblock


\bibitem[Koren et~al\mbox{.}(2009)]%
        {koren2009matrix}
\bibfield{author}{\bibinfo{person}{Yehuda Koren}, \bibinfo{person}{Robert
  Bell}, {and} \bibinfo{person}{Chris Volinsky}.}
  \bibinfo{year}{2009}\natexlab{}.
\newblock \showarticletitle{Matrix Factorization Techniques for Recommender
  Systems}.
\newblock \bibinfo{journal}{\emph{Computer}} \bibinfo{volume}{42},
  \bibinfo{number}{8} (\bibinfo{year}{2009}), \bibinfo{pages}{30--37}.
\newblock


\bibitem[Koren and Sill(2011)]%
        {koren2011OrdRec}
\bibfield{author}{\bibinfo{person}{Yehuda Koren} {and} \bibinfo{person}{Joe
  Sill}.} \bibinfo{year}{2011}\natexlab{}.
\newblock \showarticletitle{{{OrdRec}}: An Ordinal Model for Predicting
  Personalized Item Rating Distributions}. In
  \bibinfo{booktitle}{\emph{Proceedings of the Fifth {{ACM}} Conference on
  {{Recommender}} Systems}}. \bibinfo{publisher}{{ACM}},
  \bibinfo{pages}{117--124}.
\newblock


\bibitem[Lim and Teh(2007)]%
        {lim2007Variational}
\bibfield{author}{\bibinfo{person}{Yew~Jin Lim} {and} \bibinfo{person}{Yee~Whye
  Teh}.} \bibinfo{year}{2007}\natexlab{}.
\newblock \showarticletitle{Variational {{Bayesian Approach}} to {{Movie Rating
  Prediction}}}.
\newblock  (\bibinfo{year}{2007}).
\newblock


\bibitem[Marlin(2004)]%
        {marlin2004collaborative}
\bibfield{author}{\bibinfo{person}{Benjamin Marlin}.}
  \bibinfo{year}{2004}\natexlab{}.
\newblock \bibinfo{booktitle}{\emph{Collaborative filtering: A machine learning
  perspective}}.
\newblock \bibinfo{publisher}{University of Toronto Toronto}.
\newblock


\bibitem[Mazurowski(2013)]%
        {mazurowski2013Estimating}
\bibfield{author}{\bibinfo{person}{Maciej~A. Mazurowski}.}
  \bibinfo{year}{2013}\natexlab{}.
\newblock \showarticletitle{Estimating Confidence of Individual Rating
  Predictions in Collaborative Filtering Recommender Systems}.
\newblock \bibinfo{journal}{\emph{Expert Systems with Applications}}
  \bibinfo{volume}{40}, \bibinfo{number}{10} (\bibinfo{year}{2013}),
  \bibinfo{pages}{3847--3857}.
\newblock
\showISSN{0957-4174}


\bibitem[McNee et~al\mbox{.}(2003)]%
        {mcnee2003Confidence}
\bibfield{author}{\bibinfo{person}{Sean McNee}, \bibinfo{person}{Shyong Lam},
  \bibinfo{person}{Catherine Guetzlaff}, \bibinfo{person}{Joseph Konstan},
  {and} \bibinfo{person}{John Riedl}.} \bibinfo{year}{2003}\natexlab{}.
\newblock \showarticletitle{Confidence {{Displays}} and {{Training}} in
  {{Recommender Systems}}.}
\newblock


\bibitem[Mesas and Bellog{\'i}n(2020)]%
        {mesas2020Exploiting}
\bibfield{author}{\bibinfo{person}{Rus~M. Mesas} {and}
  \bibinfo{person}{Alejandro Bellog{\'i}n}.} \bibinfo{year}{2020}\natexlab{}.
\newblock \showarticletitle{Exploiting Recommendation Confidence in
  Decision-Aware Recommender Systems}.
\newblock \bibinfo{journal}{\emph{Journal of Intelligent Information Systems}}
  \bibinfo{volume}{54}, \bibinfo{number}{1} (\bibinfo{year}{2020}),
  \bibinfo{pages}{45--78}.
\newblock
\showISSN{0925-9902, 1573-7675}


\bibitem[Mnih and Salakhutdinov(2007)]%
        {mnih2007Probabilistic}
\bibfield{author}{\bibinfo{person}{Andriy Mnih} {and} \bibinfo{person}{Russ~R
  Salakhutdinov}.} \bibinfo{year}{2007}\natexlab{}.
\newblock \showarticletitle{Probabilistic {{Matrix Factorization}}}. In
  \bibinfo{booktitle}{\emph{Advances in {{Neural Information Processing
  Systems}}}}, Vol.~\bibinfo{volume}{20}. \bibinfo{publisher}{{Curran
  Associates, Inc.}}
\newblock


\bibitem[Moon et~al\mbox{.}(2020)]%
        {moon2020confidence}
\bibfield{author}{\bibinfo{person}{Jooyoung Moon}, \bibinfo{person}{Jihyo Kim},
  \bibinfo{person}{Younghak Shin}, {and} \bibinfo{person}{Sangheum Hwang}.}
  \bibinfo{year}{2020}\natexlab{}.
\newblock \showarticletitle{Confidence-aware learning for deep neural
  networks}. In \bibinfo{booktitle}{\emph{international conference on machine
  learning}}. PMLR, \bibinfo{pages}{7034--7044}.
\newblock


\bibitem[Peska and Balcar(2022)]%
        {peska2022Effect}
\bibfield{author}{\bibinfo{person}{Ladislav Peska} {and}
  \bibinfo{person}{Stepan Balcar}.} \bibinfo{year}{2022}\natexlab{}.
\newblock \showarticletitle{The {{Effect}} of {{Feedback Granularity}} on
  {{Recommender Systems Performance}}}. In \bibinfo{booktitle}{\emph{Sixteenth
  {{ACM Conference}} on {{Recommender Systems}}}}. \bibinfo{publisher}{{ACM}},
  \bibinfo{pages}{586--591}.
\newblock


\bibitem[Rechkemmer and Yin(2022)]%
        {rechkemmer2022confidence}
\bibfield{author}{\bibinfo{person}{Amy Rechkemmer} {and} \bibinfo{person}{Ming
  Yin}.} \bibinfo{year}{2022}\natexlab{}.
\newblock \showarticletitle{When Confidence Meets Accuracy: Exploring the
  Effects of Multiple Performance Indicators on Trust in Machine Learning
  Models}. In \bibinfo{booktitle}{\emph{CHI Conference on Human Factors in
  Computing Systems}}. \bibinfo{pages}{1--14}.
\newblock


\bibitem[Reilly et~al\mbox{.}(2005)]%
        {reilly2005Critiquing}
\bibfield{author}{\bibinfo{person}{James Reilly}, \bibinfo{person}{Barry
  Smyth}, \bibinfo{person}{Lorraine McGinty}, {and} \bibinfo{person}{Kevin
  McCarthy}.} \bibinfo{year}{2005}\natexlab{}.
\newblock \showarticletitle{Critiquing with {{Confidence}}}. In
  \bibinfo{booktitle}{\emph{Case-{{Based Reasoning Research}} and
  {{Development}}}}, \bibfield{editor}{\bibinfo{person}{H{\'e}ctor
  {Mu{\~n}oz-{\'A}vila}} {and} \bibinfo{person}{Francesco Ricci}} (Eds.).
  \bibinfo{publisher}{{Springer}}, \bibinfo{pages}{436--450}.
\newblock


\bibitem[Rendle et~al\mbox{.}(2020)]%
        {rendle2020Neural}
\bibfield{author}{\bibinfo{person}{Steffen Rendle}, \bibinfo{person}{Walid
  Krichene}, \bibinfo{person}{Li Zhang}, {and} \bibinfo{person}{John
  Anderson}.} \bibinfo{year}{2020}\natexlab{}.
\newblock \showarticletitle{Neural {{Collaborative Filtering}} vs. {{Matrix
  Factorization Revisited}}}. In \bibinfo{booktitle}{\emph{Fourteenth {{ACM
  Conference}} on {{Recommender Systems}}}}. \bibinfo{publisher}{{ACM}},
  \bibinfo{pages}{240--248}.
\newblock


\bibitem[Rendle et~al\mbox{.}(2019)]%
        {rendle2019difficulty}
\bibfield{author}{\bibinfo{person}{Steffen Rendle}, \bibinfo{person}{Li Zhang},
  {and} \bibinfo{person}{Yehuda Koren}.} \bibinfo{year}{2019}\natexlab{}.
\newblock \showarticletitle{On the difficulty of evaluating baselines: A study
  on recommender systems}.
\newblock \bibinfo{journal}{\emph{arXiv preprint arXiv:1905.01395}}
  (\bibinfo{year}{2019}).
\newblock


\bibitem[Ricci et~al\mbox{.}(2015)]%
        {ricci2015recommender}
\bibfield{author}{\bibinfo{person}{Francesco Ricci}, \bibinfo{person}{Lior
  Rokach}, {and} \bibinfo{person}{Bracha Shapira}.}
  \bibinfo{year}{2015}\natexlab{}.
\newblock \showarticletitle{Recommender Systems: Introduction and Challenges}.
\newblock In \bibinfo{booktitle}{\emph{Recommender Systems Handbook}}.
  \bibinfo{publisher}{Springer}, \bibinfo{pages}{1--34}.
\newblock


\bibitem[{Robinson-Cox} and Boik(1998)]%
        {robinson-cox1998Derivatives}
\bibfield{author}{\bibinfo{person}{James {Robinson-Cox}} {and}
  \bibinfo{person}{Robert Boik}.} \bibinfo{year}{1998}\natexlab{}.
\newblock \showarticletitle{Derivatives of the {{Incomplete Beta Function}}}.
\newblock \bibinfo{journal}{\emph{Journal of Statistical Software}}
  \bibinfo{volume}{03} (\bibinfo{year}{1998}).
\newblock


\bibitem[Salakhutdinov and Mnih(2008)]%
        {salakhutdinov2008Bayesian}
\bibfield{author}{\bibinfo{person}{Ruslan Salakhutdinov} {and}
  \bibinfo{person}{Andriy Mnih}.} \bibinfo{year}{2008}\natexlab{}.
\newblock \showarticletitle{Bayesian Probabilistic Matrix Factorization Using
  {{Markov}} Chain {{Monte Carlo}}}. In \bibinfo{booktitle}{\emph{Proceedings
  of the 25th International Conference on {{Machine}} Learning}}.
  \bibinfo{publisher}{{Association for Computing Machinery}},
  \bibinfo{pages}{880--887}.
\newblock


\bibitem[Schnabel et~al\mbox{.}(2016)]%
        {Schnabel2016}
\bibfield{author}{\bibinfo{person}{Tobias Schnabel}, \bibinfo{person}{Adith
  Swaminathan}, \bibinfo{person}{Ashudeep Singh}, \bibinfo{person}{Navin
  Chandak}, {and} \bibinfo{person}{Thorsten Joachims}.}
  \bibinfo{year}{2016}\natexlab{}.
\newblock \showarticletitle{Recommendations As Treatments: Debiasing Learning
  and Evaluation}. In \bibinfo{booktitle}{\emph{Proceedings of the 33rd
  International Conference on International Conference on Machine Learning}}.
  \bibinfo{pages}{1670--1679}.
\newblock


\bibitem[Sedhain et~al\mbox{.}(2015)]%
        {sedhain2015AutoRec}
\bibfield{author}{\bibinfo{person}{Suvash Sedhain},
  \bibinfo{person}{Aditya~Krishna Menon}, \bibinfo{person}{Scott Sanner}, {and}
  \bibinfo{person}{Lexing Xie}.} \bibinfo{year}{2015}\natexlab{}.
\newblock \showarticletitle{{{AutoRec}}: {{Autoencoders Meet Collaborative
  Filtering}}}. In \bibinfo{booktitle}{\emph{Proceedings of the 24th
  {{International Conference}} on {{World Wide Web}}}}.
  \bibinfo{publisher}{{ACM}}, \bibinfo{pages}{111--112}.
\newblock


\bibitem[Shardanand and Maes(1995)]%
        {shardanand1995Social}
\bibfield{author}{\bibinfo{person}{Upendra Shardanand} {and}
  \bibinfo{person}{Pattie Maes}.} \bibinfo{year}{1995}\natexlab{}.
\newblock \showarticletitle{Social Information Filtering: Algorithms for
  Automating ``Word of Mouth''}. In \bibinfo{booktitle}{\emph{Proceedings of
  the {{SIGCHI}} Conference on {{Human}} Factors in Computing Systems - {{CHI}}
  '95}}. \bibinfo{publisher}{{ACM Press}}, \bibinfo{pages}{210--217}.
\newblock


\bibitem[Steck(2013)]%
        {steck2013evaluation}
\bibfield{author}{\bibinfo{person}{Harald Steck}.}
  \bibinfo{year}{2013}\natexlab{}.
\newblock \showarticletitle{Evaluation of Recommendations: Rating-Prediction
  and Ranking}. In \bibinfo{booktitle}{\emph{Proceedings of the Seventh ACM
  Conference on Recommender Systems}}. \bibinfo{pages}{213--220}.
\newblock


\bibitem[Swaminathan and Joachims(2015)]%
        {swaminathan2015Self}
\bibfield{author}{\bibinfo{person}{Adith Swaminathan} {and}
  \bibinfo{person}{Thorsten Joachims}.} \bibinfo{year}{2015}\natexlab{}.
\newblock \showarticletitle{The {{Self-Normalized Estimator}} for
  {{Counterfactual Learning}}}. In \bibinfo{booktitle}{\emph{Advances in
  {{Neural Information Processing Systems}}}}, Vol.~\bibinfo{volume}{28}.
  \bibinfo{pages}{3231--3239}.
\newblock


\bibitem[Thompson and Barnett(1986)]%
        {thompson1986Coulomb}
\bibfield{author}{\bibinfo{person}{I.J. Thompson} {and} \bibinfo{person}{A.R.
  Barnett}.} \bibinfo{year}{1986}\natexlab{}.
\newblock \showarticletitle{Coulomb and {{Bessel}} Functions of Complex
  Arguments and Order}.
\newblock \bibinfo{journal}{\emph{J. Comput. Phys.}} \bibinfo{volume}{64},
  \bibinfo{number}{2} (\bibinfo{year}{1986}), \bibinfo{pages}{490--509}.
\newblock
\showISSN{00219991}


\bibitem[Wang et~al\mbox{.}(2018)]%
        {wang2018ConfidenceAware}
\bibfield{author}{\bibinfo{person}{Chao Wang}, \bibinfo{person}{Qi Liu},
  \bibinfo{person}{Runze Wu}, \bibinfo{person}{Enhong Chen},
  \bibinfo{person}{Chuanren Liu}, \bibinfo{person}{Xunpeng Huang}, {and}
  \bibinfo{person}{Zhenya Huang}.} \bibinfo{year}{2018}\natexlab{}.
\newblock \showarticletitle{Confidence-{{Aware Matrix Factorization}} for
  {{Recommender Systems}}}.
\newblock \bibinfo{journal}{\emph{Proceedings of the AAAI Conference on
  Artificial Intelligence}} \bibinfo{volume}{32}, \bibinfo{number}{1}
  (\bibinfo{year}{2018}).
\newblock
\showISSN{2374-3468, 2159-5399}


\bibitem[Zhang et~al\mbox{.}(2016)]%
        {zhang2016Prediction}
\bibfield{author}{\bibinfo{person}{Mingyue Zhang}, \bibinfo{person}{Xunhua
  Guo}, {and} \bibinfo{person}{Guoqing Chen}.} \bibinfo{year}{2016}\natexlab{}.
\newblock \showarticletitle{Prediction Uncertainty in Collaborative Filtering:
  {{Enhancing}} Personalized Online Product Ranking}.
\newblock \bibinfo{journal}{\emph{Decision Support Systems}}
  \bibinfo{volume}{83} (\bibinfo{year}{2016}), \bibinfo{pages}{10--21}.
\newblock
\showISSN{0167-9236}


\bibitem[Zhang et~al\mbox{.}(2020)]%
        {zhang2020explainable}
\bibfield{author}{\bibinfo{person}{Yongfeng Zhang}, \bibinfo{person}{Xu Chen},
  {et~al\mbox{.}}} \bibinfo{year}{2020}\natexlab{}.
\newblock \showarticletitle{Explainable recommendation: A survey and new
  perspectives}.
\newblock \bibinfo{journal}{\emph{Foundations and Trends{\textregistered} in
  Information Retrieval}} \bibinfo{volume}{14}, \bibinfo{number}{1}
  (\bibinfo{year}{2020}), \bibinfo{pages}{1--101}.
\newblock


\bibitem[Zhao et~al\mbox{.}(2015)]%
        {zhao2015Bayesian}
\bibfield{author}{\bibinfo{person}{Qibin Zhao}, \bibinfo{person}{Liqing Zhang},
  {and} \bibinfo{person}{Andrzej Cichocki}.} \bibinfo{year}{2015}\natexlab{}.
\newblock \showarticletitle{Bayesian {{CP Factorization}} of {{Incomplete
  Tensors}} with {{Automatic Rank Determination}}}.
\newblock \bibinfo{journal}{\emph{IEEE Transactions on Pattern Analysis and
  Machine Intelligence}} \bibinfo{volume}{37}, \bibinfo{number}{9}
  (\bibinfo{year}{2015}), \bibinfo{pages}{1751--1763}.
\newblock
\showISSN{1939-3539}


\end{thebibliography}

\end{document}